\documentclass[a4paper,fleqn]{cas-sc}

\usepackage[numbers]{natbib}
\usepackage[capitalise]{cleveref}
\usepackage{subcaption}
\usepackage{multirow}
\usepackage{tikz}

\def\tsc#1{\csdef{#1}{\textsc{\lowercase{#1}}\xspace}}
\tsc{WGM}
\tsc{QE}

\begin{document}

\let\WriteBookmarks\relax
\def\floatpagepagefraction{1}
\def\textpagefraction{.001}

\shorttitle{Auto Window}

\shortauthors{Yiqin, et~al.}

\title [mode = title]{
    Interpretable Auto Window Setting for Deep-Learning-Based CT Analysis
}



\author[1]{Yiqin Zhang}[orcid=0000-0003-2099-2687]
\cormark[1]
\ead{zyqmgam@163.com}
\ead[url]{https://github.com/MGAMZ}
\credit{Project administration, Conceptualization, Data curation, Formal analysis, Investigation, Methodology, Software, Resources, Validation, Visualization, Writing - Original draft preparation, Writing - review and editing}
\cortext[1]{Corresponding author}

\author[1]{Meiling Chen}
\ead{meiling_chen0313@163.com}
\credit{Formal Analysis, Validation, Visualization, Writing - Reviewing and Editing}

\author[2]{Zhengjie Zhang}
\ead{18715483478@163.com}
\credit{Data curation, Formal Analysis, Investigation}

\affiliation[1]{
    organization={University of Shanghai for Science and Technology},
    city={Shanghai},
    country={China}}
\affiliation[2]{
    organization={Huashan Hospital, Fudan University},
    city={Shanghai},
    country={China}}


\begin{abstract}
    Whether during the early days of popularization or in the present, the window setting in Computed Tomography (CT) has always been an indispensable part of the CT analysis process. Although research has investigated the capabilities of CT multi-window fusion in enhancing neural networks, there remains a paucity of domain-invariant, intuitively interpretable methodologies for Auto Window Setting. In this work, we propose an plug-and-play module originate from \textit{Tanh} activation function, which is compatible with mainstream deep learning architectures. Starting from the physical principles of CT, we adhere to the principle of interpretability to ensure the module's reliability for medical implementations. The domain-invariant design facilitates observation of the preference decisions rendered by the adaptive mechanism from a clinically intuitive perspective. This enables the proposed method to be understood not only by experts in neural networks but also garners higher trust from clinicians. We confirm the effectiveness of the proposed method in multiple open-source datasets, yielding $10\%\sim200\%+$ Dice improvements on hard segment targets.
\end{abstract}

\begin{graphicalabstract}
\includegraphics[width=\linewidth]{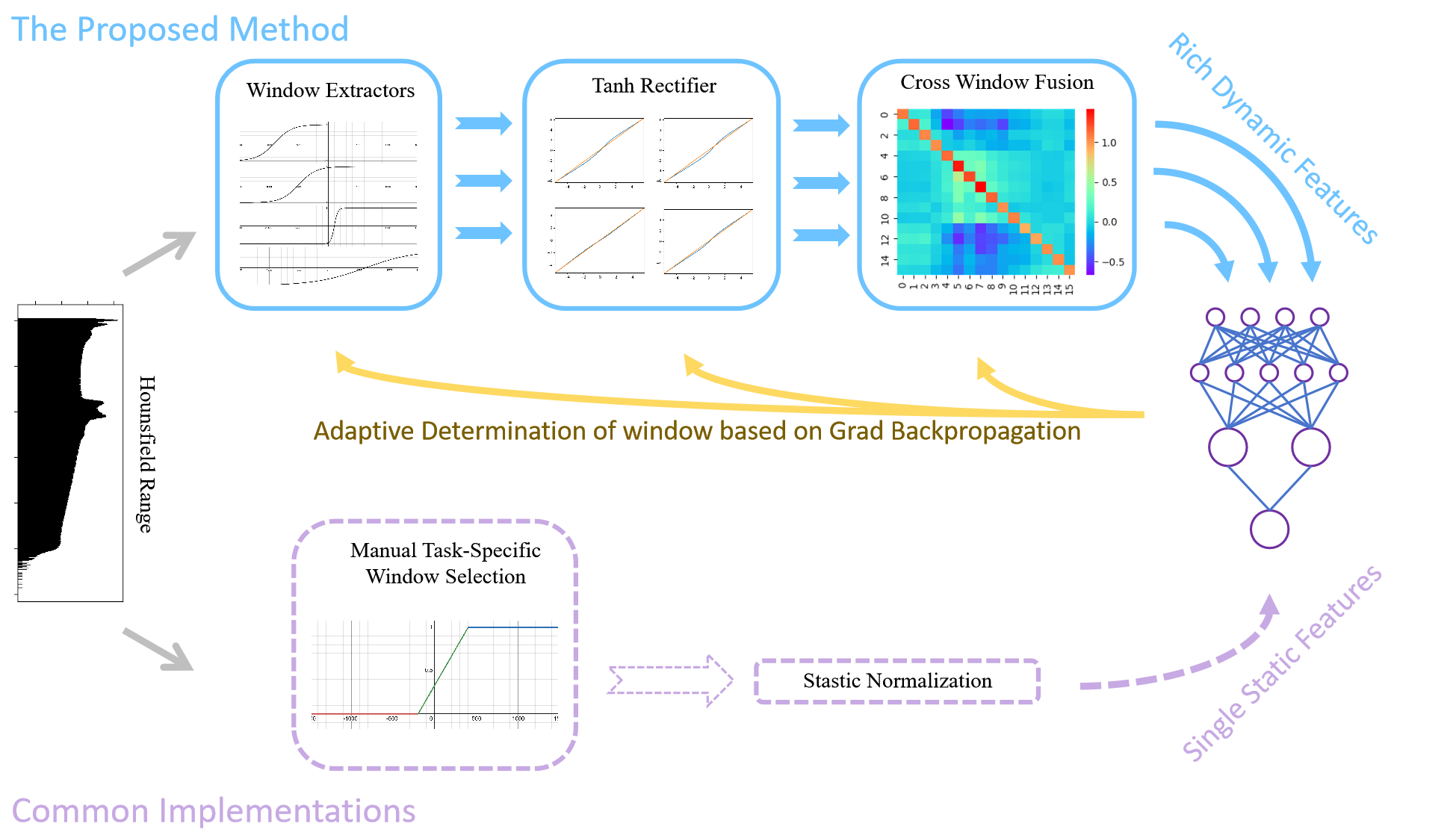}
\end{graphicalabstract}

\begin{highlights}
\item A plugin module for neural networks to automatically set CT windows.
\item Interpretable and compatible to mainstream deep learning architectures.
\item Being able to help researchers and developers to construct deep-learning-based frameworks more easily.
\item Refined domain-invariant design is used to enhance the controllability and interpretability of the module.
\end{highlights}

\begin{keywords}
\sep Deep Learning \sep Medical Image Analysis \sep Computed Tomography \sep Multi-Window Processing \sep Medical Fundamental Models
\end{keywords}

\maketitle
\setlength{\parskip}{\baselineskip}

\section{Introduction and Related Works}

\subsection{Recent Advances in Medical CT Analysis}

In recent years, the society has witnessed the explosive growth of deep learning and its widespread application in the medical field \citep{Gheisari2023, SHAMSHAD2023}. One important application is the use of visual neural networks for automated analysis of radiological images to assist doctors in efficiently diagnosing various diseases. In this topic, several subfields are currently hot research focuses, such as registration, segmentation, classification, reconstruction, semi-supervision, class imbalance, etc.. These visual models consist of conceptual components such as "Embedding, Backbone, Task Head." Embedding is responsible for "translating" raw image data into a high-dimensional implicit feature that neural networks can process, the Backbone is responsible for analyzing the feature, and the Task Head generates the required output. The majority researches use Unet-Style neural network architecture, which is a widely accepted and effective architecture for medical image analysis and universal visual segmentation tasks \citep{Unet, VNet}.

\subsection{Window: An Essential Component}

\begin{figure}[tbp]
    \centering
    \begin{subfigure}[b]{0.19\linewidth}
        \includegraphics[width=\linewidth]{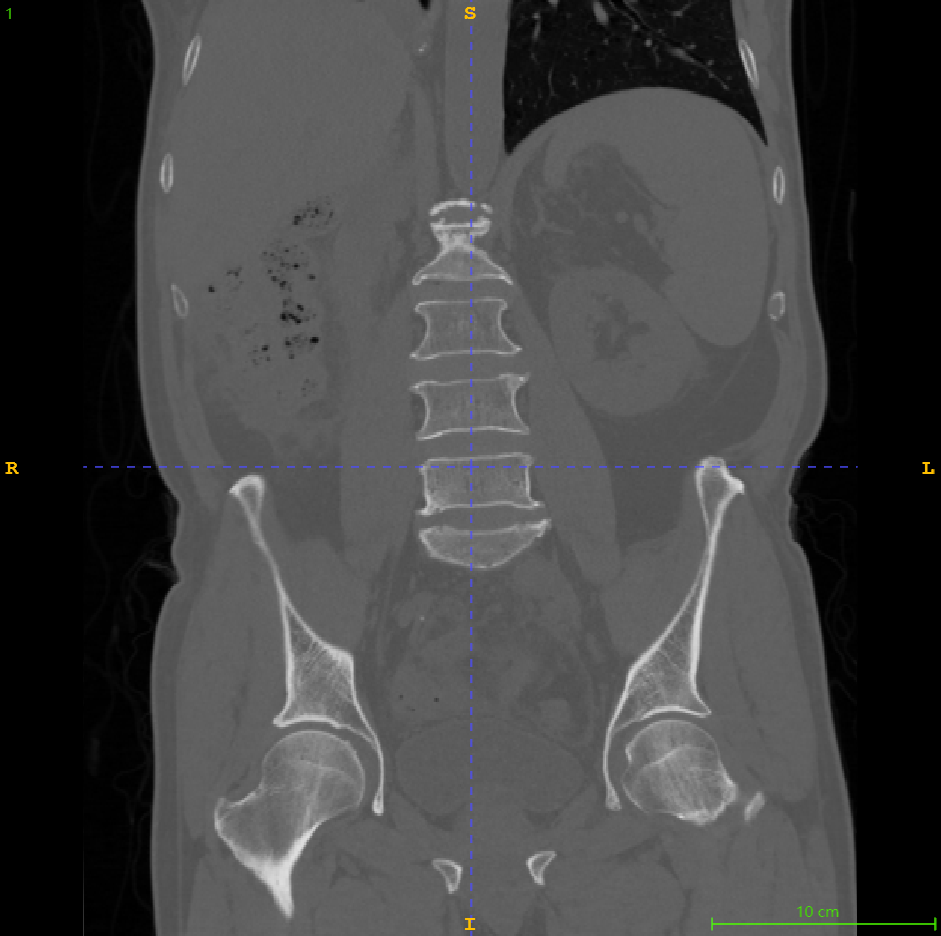}
        \caption{Whole Window}
    \end{subfigure}
    \begin{subfigure}[b]{0.19\linewidth}
        \includegraphics[width=\linewidth]{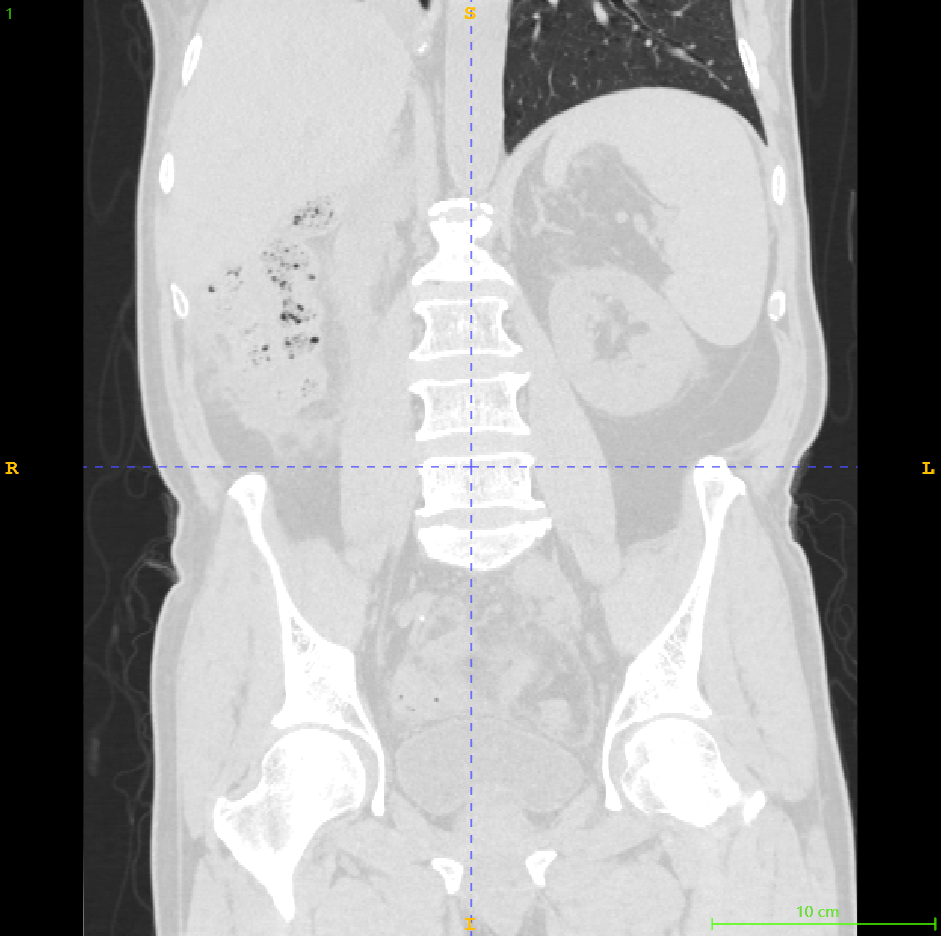}
        \caption{Lung Window}
    \end{subfigure}
    \begin{subfigure}[b]{0.19\linewidth}
        \includegraphics[width=\linewidth]{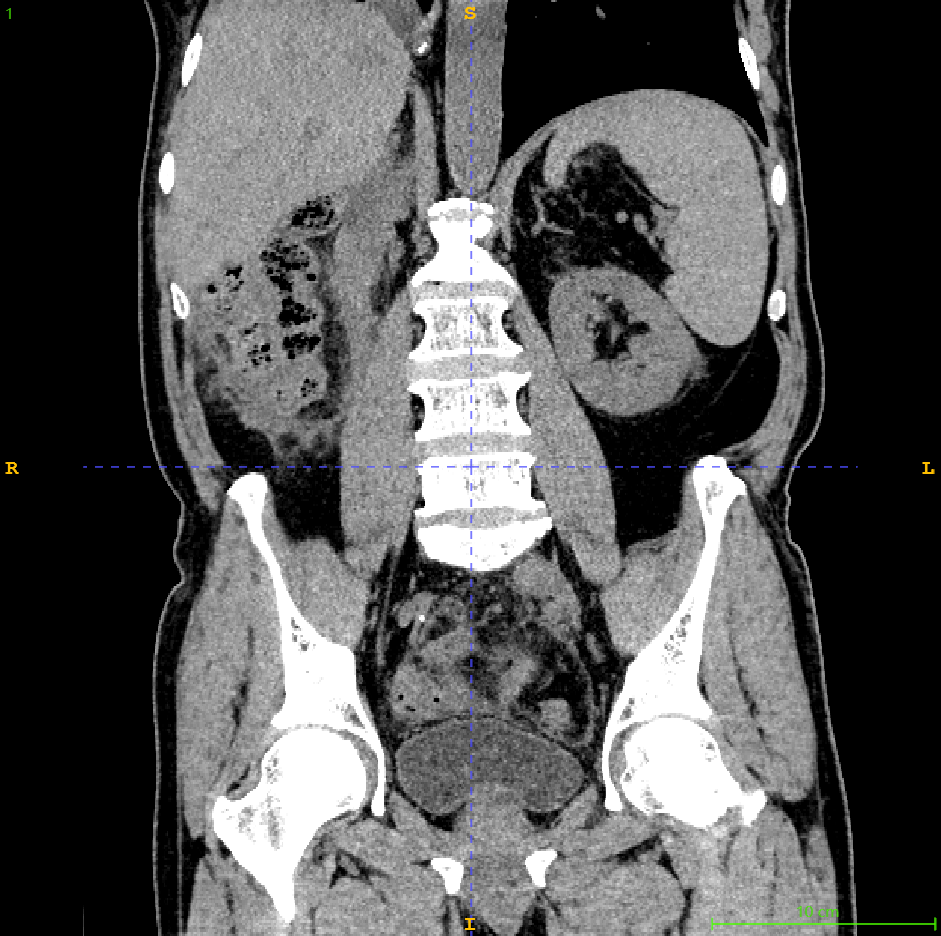}
        \caption{Medias. Window}
    \end{subfigure}
    \begin{subfigure}[b]{0.19\linewidth}
        \includegraphics[width=\linewidth]{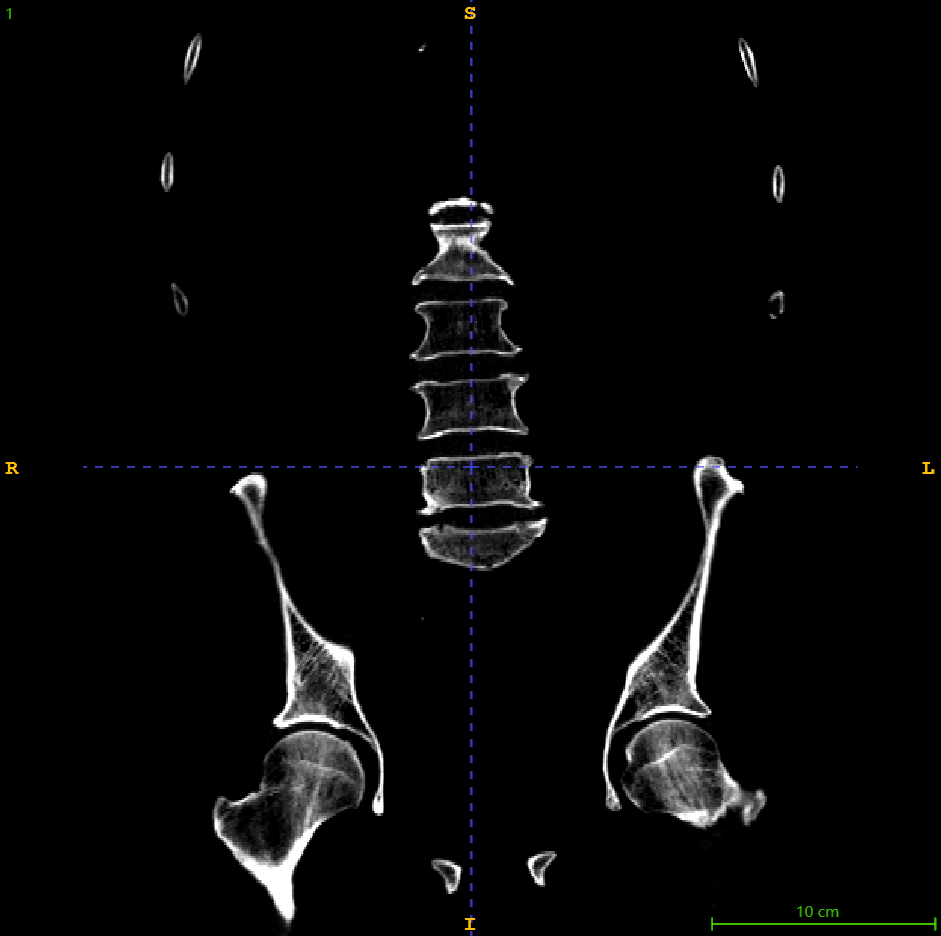}
        \caption{Bone Window}
    \end{subfigure}
    \begin{subfigure}[b]{0.19\linewidth}
        \includegraphics[width=\linewidth]{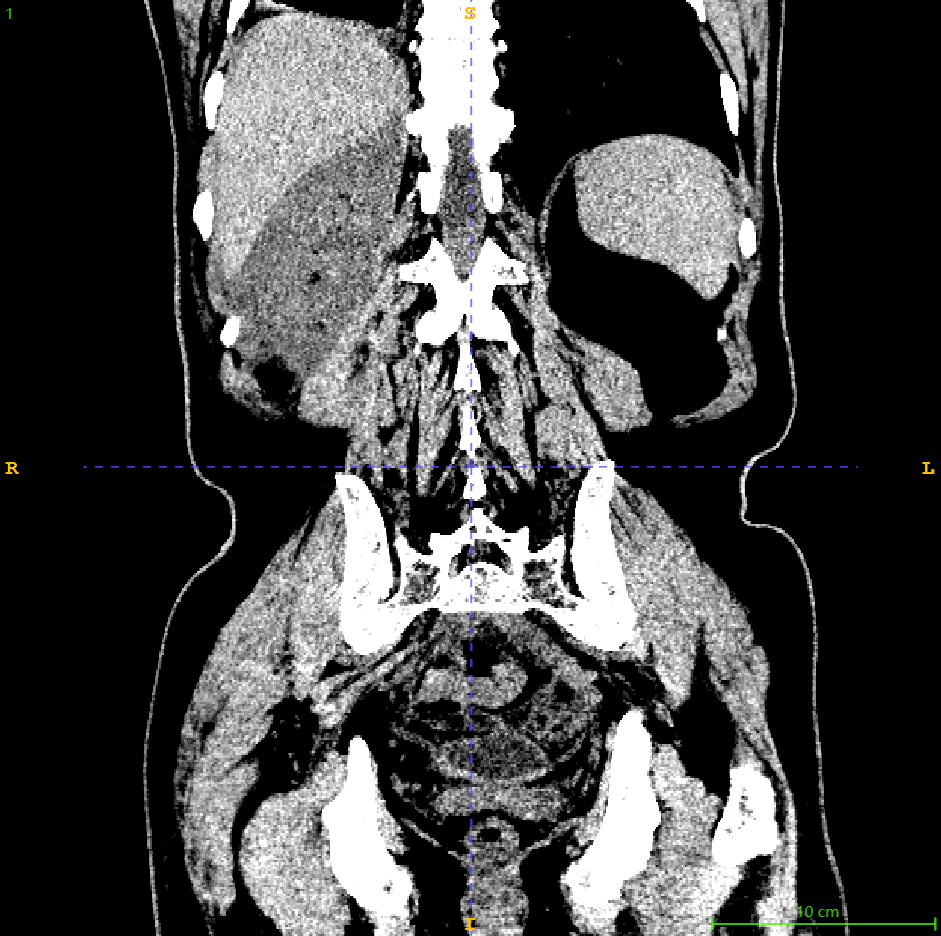}
        \caption{Liver Window}
    \end{subfigure}
    \caption{Different Window Settings. For different tasks, it is often necessary to first select an appropriate window before proceeding with subsequent preprocessing and feature recognition operations. In clinical practice, doctors also frequently choose specific windows to diagnose certain symptoms or diseases.}
    \label{fig:DiffWindowSetting}
\end{figure}

Radiologists also go through a process similar to "translation" when interpreting CT scans - Window Setting (\cref{fig:DiffWindowSetting}). It has a long history and has been highly regarded since its proposal. Currently, this step is a necessity when performing most CT analyses. Window setting can extract the interested Hounsfield Unit (HU) range \citep{DenOtterHounsfield} from the original CT image, allowing radiologists to observe certain organs with clarity and precision. Due to the much larger dynamic range of CT compared to conventional digital imaging, skipping this step would result in CT images that cannot distinguish tissue boundaries or lesion locations when observed by the human eye. Nowadays, the role of medical visual neural networks is to replace the human eye and generate judgments; therefore, the vast majority of neural networks prefers voxels that has undergone window setting just like humans. 

In studies focusing on the lungs, \citet{Chen10093869} allege that the CT window selection is vital on lung nodule detection and irreversibly discarded image semantics below -1000 or above 400 HU, and achieved good precision in nodule detection; \citet{Mascalchi2017} included relative area (RA) at -970, -960 or -950 HU to measure lung density. And in studies focusing on the bones, \citet{Yu10137862} used $[-100, 400]$ HU to implement cortical bone separation task; \citet{Prakhar10.1117} used $[350, 1000]$ HU to segment bone from whole-body CT scan.

In other studies, \citet{Wen978} limited HU according to the slice area corresponding to the annotations to perform liver tumor and vessel segmentation; \citet{Ruiquan10} directly predefine the HU values corresponding to blood clots as 60-80 to assist the model in identifying cerebral hemorrhage events; \citet{Chen10635129} used $50\pm175$ HU when training their gastric tumor segmentation model; \citet{YanDifferentiation} concluded that HU$\leq42.5$ should be treated as a critical element when analyzing gastrointestinal stromal tumors (GISTs).

Further, \citet{AbboudImpactWindow} alleged the window setting is still a manual process within highly-automatic deep-learning-based methods, which is not only task-specific but also requires some professional knowledge. \citet{BELLENS2024324} pointed out that nowadays' faster acquisition protocols complicates the segmentation step, the configurations during CT Image analysis require manual adjustment. \citet{Cruz2019} also pointed out that the HU value will be influenced by different post-processing methods, e.g. reconstruction and denoising. These studies all suggested: when the method can dynamically handle data from different HUs, it has superior application potential and stronger generalizability in clinical settings. Moreover, when handling large-scale multi-target segmentation or detection tasks, the window setting can be knotty to achieve acceptable performance on all targets. In which case, the researchers often fall to the compromise of using whole window or wide window.

\subsection{More Efficient Window Setting}

Based on the above observations, \citet{Kwon2020} proposed a learnable setting method for liver CT segmentation task; \citet{KARKI2020101850} proposed a similar method for hemorrhage detection task, but still needs initial window assignment; \citet{Yuankai2019} used a random solution to combine multiple window levels, thus providing richer feature for neural network; \citet{Han2024} used multi window to mitigate the long-tail effect in CT diagnostic data for Covid-19; \citet{HOOGI201746} introduced an iterative and adaptive point-wise window setting method, achieved great accuracy, but required significant higher computational overhead and deployment complexity; \citet{jimaging9120270} introduced multi-learnable window on chest X-ray classification task, the window setting operation is placed after the embedding layer. The majority of the current works lack interpretability \citep{Goceri2023}, as the processed value is neither directly related to the HU domain nor easy to visualize the learned projection rules. In medical domains, clinical practitioners often prefers a more interpretable method, given that black-box models always make it difficult for doctors to understand the "Why" \citep{technologies9010002, Frasca2024}.

\subsection{Our Contributions}

In this work, we proposed an automated learnable method that can replace manual window setting. It can be added to existing neural networks in a manner similar to an Embedding module, automatically determining the range of interested HU sub-domain based on specific downstream tasks. This means that when using neural networks for CT image analysis, there is no longer a need for manual window setting. This automated approach is not only robust but also more likely to achieve superior results, as manual parameter settings often rely on empirical values and may vary across different scanning and post-processing protocols.

Compared to the existing methods, ours has the following several advantages: 
\begin{itemize}
    \item \textbf{Domain-Invarient Design.} The method proposed is characterized by a limited number of controllable adaptive parameters and utilizes a mathematically defined model to construct a domain-invariant dynamic mapping function. The absence of deep neural networks in this approach ensures that the model’s outputs remain interpretable for clinical professionals.
    \item \textbf{Module-Wise Independency.} The proposed method consists of three sub-modules, each of which can be independently analyzed and optimized, thus ensuring their interpretability. The mapping process can still be executed smoothly even when one or two modules are missing.
    \item \textbf{Progressive Nonlinearity.} The approach avoids reliance on a single, complex black-box model with an extensive set of parameters for mapping the HU domain. Instead, it incrementally enhances the nonlinearity of the mapping process in distinct stages. This strategy not only provides users with greater flexibility in model adjustment but also facilitates clearer analytical and observational insights.
\end{itemize}

The subsequent sections are structured as follows:

\cref{sec:Prelim} delineates the overarching design framework of the method proposed by us and conducts a feasibility analysis thereof.

\cref{sec:Method} comprises three distinct parts, each dedicated to a sub-module of the proposed method, examined through mathematical and design frameworks: Adaptive Window Extractor (\cref{sec:Method_WinE}), the Tanh-Based Post Rectificator (\cref{sec:Method_TRec}), and the Paralleled Windows and Fusion mechanisms (\cref{sec:Method_CrsF}).

In \cref{sec:Exp}, we commence by presenting the dataset and the fundamental experimental parameters (\cref{sec:Exp_Dataset}). This is followed by an evaluation of our method’s end-to-end performance on large (\cref{sec:Exp_LargeScale}) and small (\cref{sec:Exp_SmallScale}) scale datasets. \cref{sec:Exp_Interp} is allocated to a comprehensive examination and validation of the interpretability of the three proposed sub-modules. \cref{sec:Exp_Perf} concludes with a concise analysis of the computational overhead.

In the concluding \cref{sec:Conc}, we offer a synthesis of our findings and propose avenues for future research.

\section{Preliminaries}
\label{sec:Prelim}

\subsection{Definition of the Proposed Module}

Currently, data-driven medical imaging analysis models primarily utilize: \textbf{1)} dataset $\mathcal{S}$, \textbf{2)} data cleaning and preprocessing $\mathcal{F}$, \textbf{3)} neural networks $\mathbf{N}$ with a large number of learnable parameters, and \textbf{4)} output or visualization $\mathcal{O}$ to generate task-specific outputs. Further, most neural-network-based methods incorporate an embedding layer $\mathbf{E}$ to as an initial projection from source domain to hidden domains. The proposed module $\mathbf{P}$ is designed to be inserted between $\mathcal{F}$ and $\mathbf{E}$, that is to say, a pre-Embedding layer. This design allows $\mathbf{N}$ to be easily integrated into the vast majority of existing neural network implementations. After incorporating the method proposed in this paper, the abstract mathematical process is outlined as follows:
\begin{equation}
    \mathbf{E}(\mathbf{P}(\mathcal{F}(\mathcal{S}))) \rightarrow \mathbf{N} \rightarrow \mathcal{O}
\end{equation}

\begin{figure}[tbp]
    \centering
    \includegraphics[width=\linewidth]{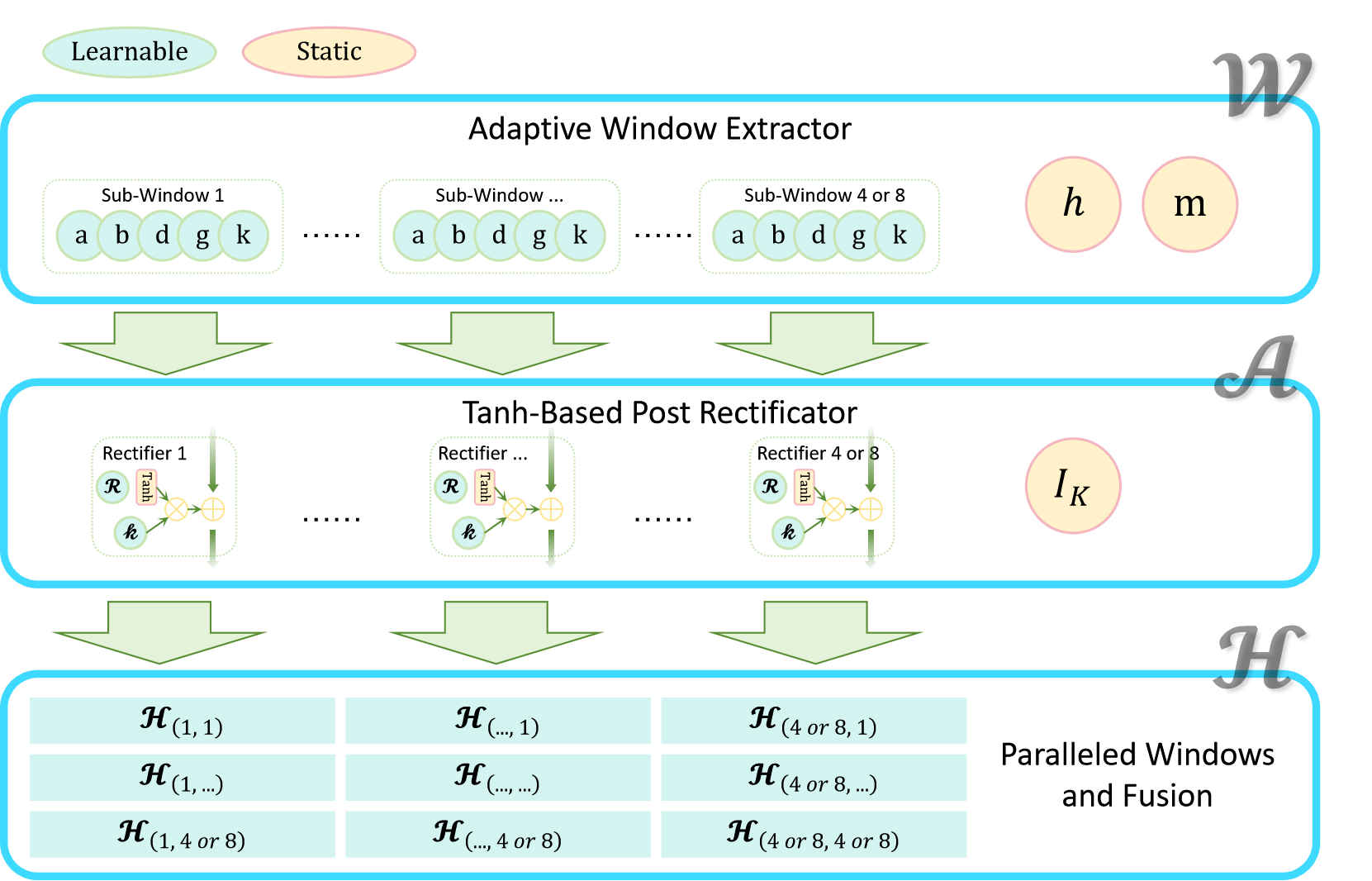}
    \caption{Structure of the Proposed Method. Consistent with the description in the text, each submodule is represented by a symbol, displayed in the top right corner of the module. Within each module, learnable parameters are shown in green and non-learnable parameters in orange.}
    \label{fig:Module_Overview}
\end{figure}

In terms of the proposed $\mathbf{P}$, there are majorly three parts: \textbf{1)} Window Extractor $\mathcal{W}$ (\cref{sec:Method_WinE}), \textbf{2)} \textit{Tanh}-based Rectificator $\mathcal{A}$ (\cref{sec:Method_TRec}), and \textbf{3)} Paralleled-Cross-Window Stream and Fusion $\mathbf{H}$ (\cref{sec:Method_CrsF}). In \cref{fig:Module_Overview}, we have depicted the top-level module data flow design. Every symbol within the figure is directly aligned with the descriptions that follow in the text.

\subsection{Fesibility Investigation based on CT Physics}

During the CT scan process, the X-rays pass through the human body and are partially absorbed. By measuring the degree of absorption of the rays, the density distribution $\mathcal{P}$ of the tissues can be determined \citep{Seeram2010}. Current technology can achieve a binary precision of 12 bits for $\mathcal{P}$, which has already exceeded the perceptual precision of mainstream neural networks $\mathbf{N}$.

Although the perceptual capacity for a single set is limited, the embedding layer of neural networks typically does not restrict the number of input channels \cref{eq:embed_prelim}. By using a function $\mathcal{P}$ to sample a high-precision set into multiple lower-precision subsets and feeding them to the embedding layer in parallel, the subsequent neural network can fully leverage its feature-capturing capabilities.
\begin{equation}
    \label{eq:embed_prelim}
    \mathcal{X} = \mathcal{X}_1, \mathcal{X}_2, \ldots, \mathcal{X}_m  = \mathbf{E}(p_1, p_2, \ldots, p_n \mid p_i \in \mathcal{P}(\mathcal{S}))
\end{equation}

\subsection{Compatibility Investigation}

In the current design, the embedding layer of the neural network is primarily used to map the source space $\mathbb{S}$ to the neural hidden space $\mathbb{H}$. To ensure maximum compatibility, any design preceding the embedding layer should not significantly alter the numerical domain of the source space. Taking CT imaging as an example: a strong signal consistently indicates that the corresponding spatial location $v(x,y,z)$ has a higher tissue density $\mathcal{P}(x,y,z)$.

Benefit from the domain-invariant design, we will intuitively observe in the experimental section how the CT signals are enhanced. This can provide interpretability for potential practical clinical applications, which is quite valuable for the scalability of this work.

\section{Proposed Method}
\label{sec:Method}

\subsection{Adaptive Window Extractor}
\label{sec:Method_WinE}

\begin{figure}[tbp]
    \centering
    \begin{subfigure}[t]{0.11\linewidth}
        \includegraphics[width=\linewidth]{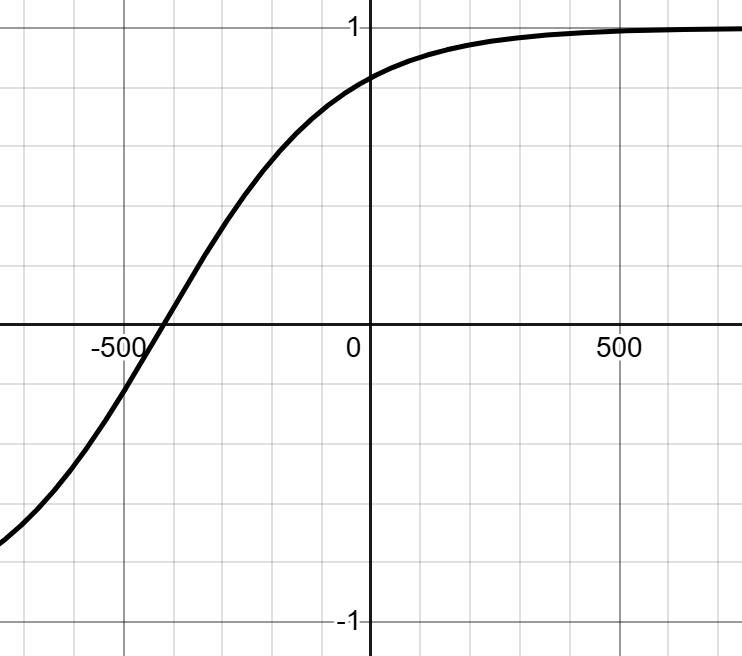}
        \caption{$a=10$}
        \label{fig:ImpactParameter_a10}
    \end{subfigure}
    \begin{subfigure}[t]{0.11\linewidth}
        \includegraphics[width=\linewidth]{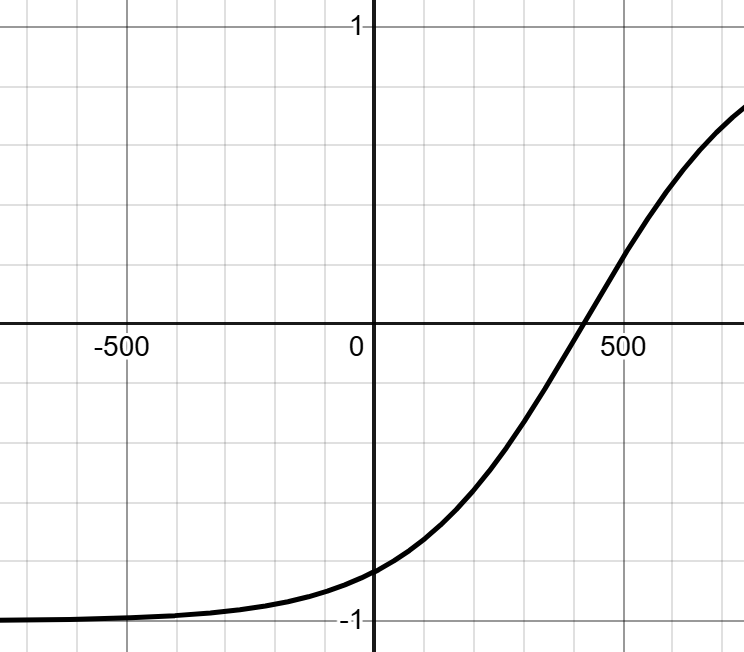}
        \caption{$b=10$}
        \label{fig:ImpactParameter_b10}
    \end{subfigure}
    \begin{subfigure}[t]{0.11\linewidth}
        \includegraphics[width=\linewidth]{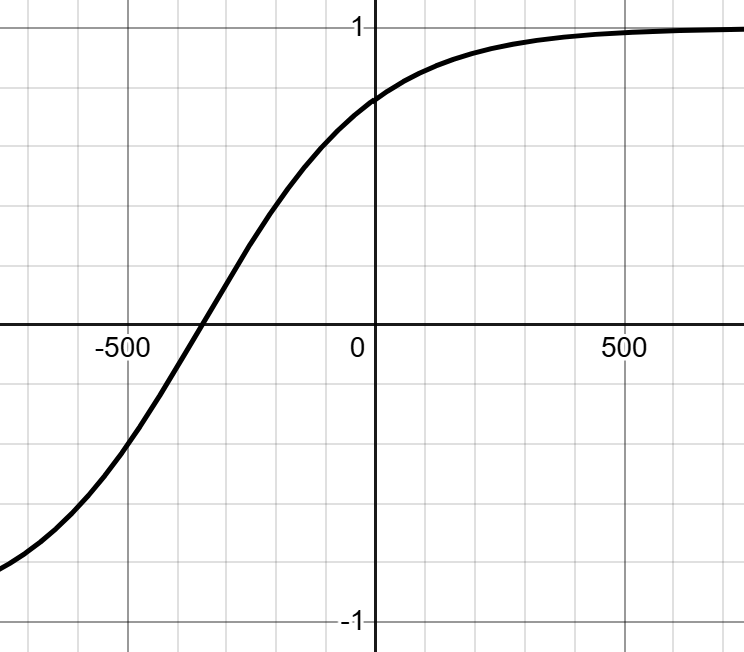}
        \caption{$d=1$}
        \label{fig:ImpactParameter_d1}
    \end{subfigure}
    \begin{subfigure}[t]{0.11\linewidth}
        \includegraphics[width=\linewidth]{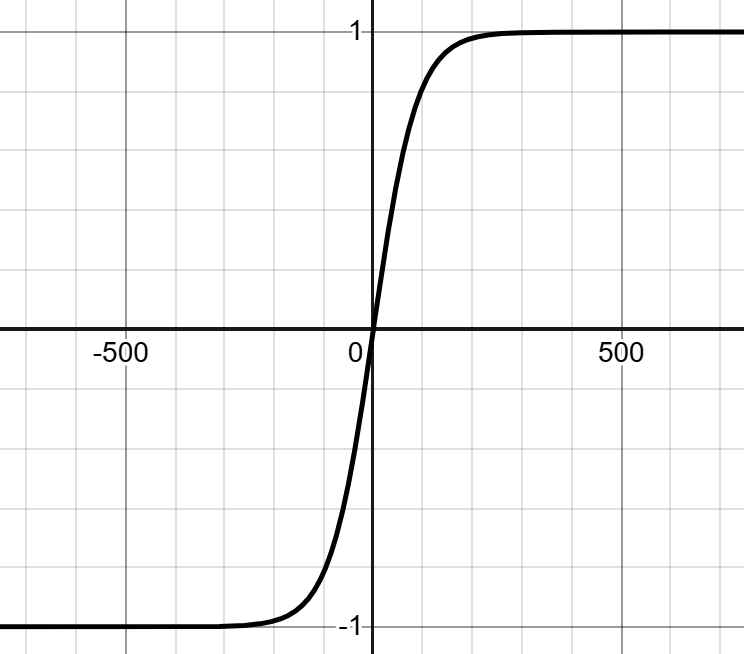}
        \caption{$g=4$}
        \label{fig:ImpactParameter_g4}
    \end{subfigure}
    \begin{subfigure}[t]{0.11\linewidth}
        \includegraphics[width=\linewidth]{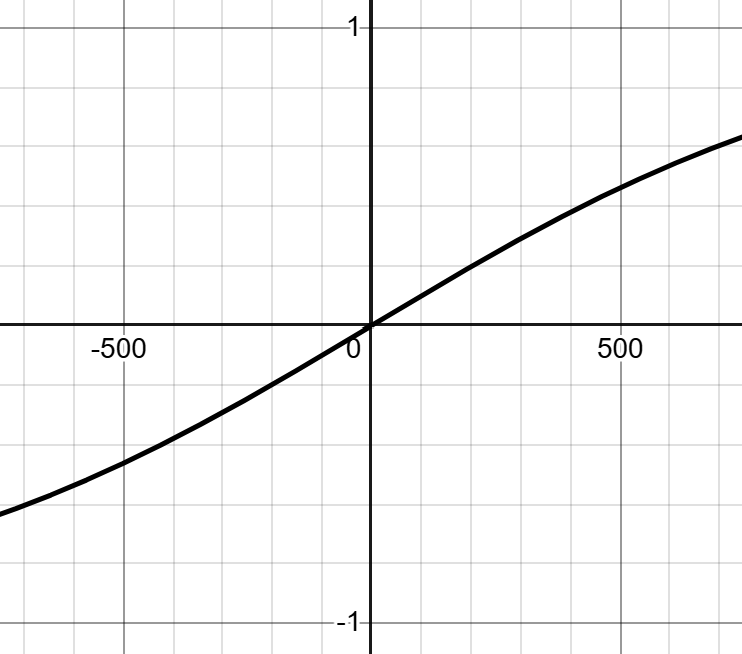}
        \caption{$h=1000$}
        \label{fig:ImpactParameter_h1000}
    \end{subfigure}
    \begin{subfigure}[t]{0.11\linewidth}
        \includegraphics[width=\linewidth]{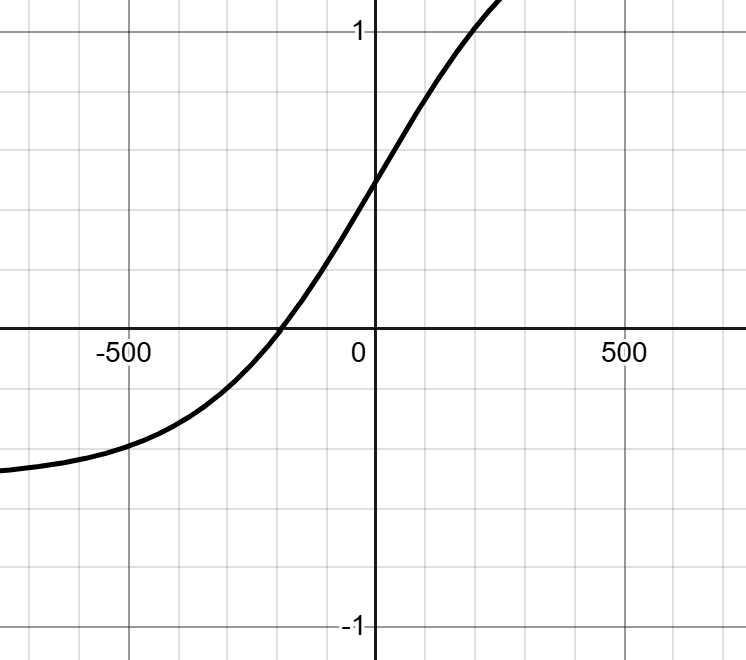}
        \caption{$k=0.5$}
        \label{fig:ImpactParameter_k0.5}
    \end{subfigure}
    \begin{subfigure}[t]{0.11\linewidth}
        \includegraphics[width=\linewidth]{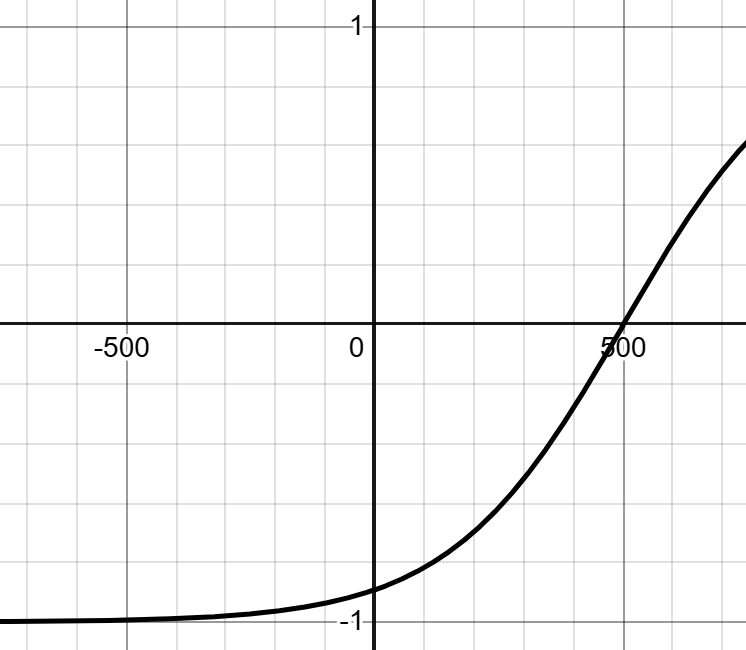}
        \caption{$m=500$}
        \label{fig:ImpactParameter_m500}
    \end{subfigure}
    \begin{subfigure}[t]{0.11\linewidth}
        \includegraphics[width=\linewidth]{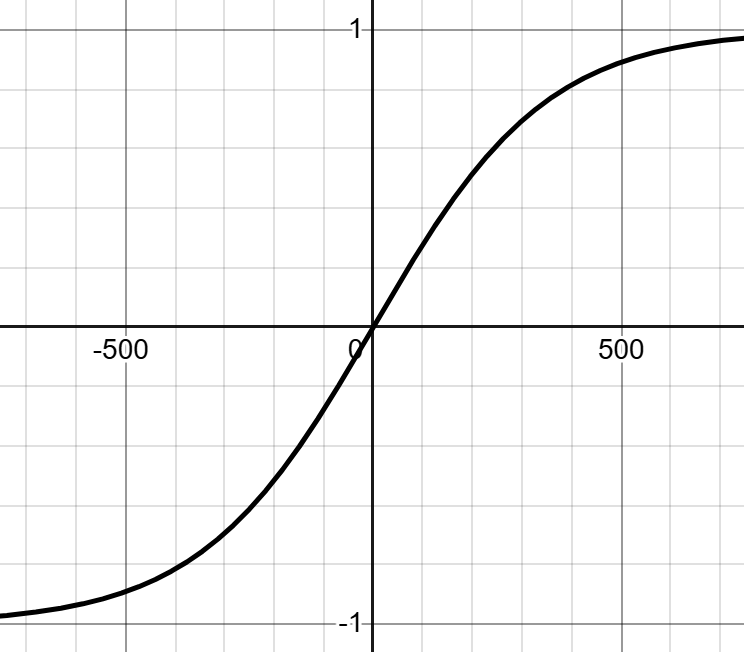}
        \caption{Baseline}
        \label{fig:ImpactParameter_baseline}
    \end{subfigure}

    \caption{Impact of all proposed parameters in Window Extractor Module $\mathcal{W}$. All other subgraphs are based on \cref{fig:ImpactParameter_baseline}, altering one parameter to visualize its role in the extraction. Baseline: $a=0, b=0, d=0, g=1, h=350, k=0, m=0$.}
    \label{fig:ImpactParameter}
\end{figure}

In whole, after obtaining the reconstructed volume $\mathcal{S}$ from the DICOM-standard CT data, we use the Window Extractor $\mathcal{W}$ as the first and most critical mapping without any preprocessing. This is a learnable normalization function derived from the \textit{Tanh} activation. It extends the reach of gradient propagation in neural network training to the initial stages of data processing. This enables the system to automatically determine the most optimal window settings, guided by gradients that are informed by the training process and predefined objectives. The mathematical process is outlined in \cref{eq:tauS} and \cref{eq:window_extractor}. The major effect provided by each parameter is shown in \cref{fig:ImpactParameter}.

\subsubsection{Originate from Tanh Activation Offset}
\label{sec:WinE_OriTanh}

In the simplest case, the normalization can be achieved using the Tanh function to directly activate the original HU domain $\mathcal{S}$, and can yield a whole-window $\mathcal{W}_{whole}(\mathcal{S}) \in (-1,1)$. All numerical features will be retained, but the nonlinear characteristics of the Tanh function determine that values near zero will occupy a larger expressive range after activation. This will also lead to an amplification of the object within the corresponding volume area, which in this case is water. As previously mentioned, clinical needs do not always require emphasizing tissues with densities close to water in CT scans. 

To achieve this, we adopt the following approach: $\mathcal{S}$ should be mapped once before the Tanh activation to shift its distribution, thereby adjusting the window level where its maximum sensitivity is located. On one hand, $\mathcal{W}(\tau(\mathcal{S})), \tau(\mathcal{S}) = \mathcal{S}-\alpha$, this introduces the simplest input offset for signal response, which can change the window location. On the other hand, $\mathcal{W}(\tau(\mathcal{S})), \tau(\mathcal{S}) = \beta \cdot \mathcal{S}$, this controls the intensity scaling, which can change the window width. This is because $\frac{d \textit{tanh}(\mathcal{S})}{d \mathcal{S}}$ is negatively correlated with $\left\lvert \mathcal{S} \right\rvert $. A larger $\beta$ ensures that only a smaller sub-source domain is close to zero after mapping, meaning that only a narrow domain is fully expressed.

Combining the two simple designs, using $\tau(\mathcal{S}) = \beta \cdot (\mathcal{S} - \alpha)$ allows us to control window location and intensity at the same time. When $\alpha$ and $\beta$ are learnable, the neural network can adjust them to achieve higher task accuracy. It is evident that this form of projection does not contain nonlinear components; the method proposed above is equivalent to adding a fully connected layer at the beginning of the model. Given that current neural networks have much more complex neural connection structures, it is difficult for a single newly added layer to bring further improvements to the whole. This is because other parts of the model have enough nonlinear degrees of freedom to easily achieve the same numerical mapping effects.

\subsubsection{Refined Offset Projection with Medical CT Prior}

Based on the characteristics of CT imaging, we introduce more prior design into this simple projection process. 

\textbf{First,} we use $\alpha = m - h \times d$, where $m$ is the major focusing, $h$ is range rectification coefficient and $d$ is dynamic interested field. $m$ is designed to directly provide window location offset to the projection. During initialization, multiple parallel windows evenly assign $m$, allowing the Window Extractor to cover all domains under initial parameters, preventing the training from starting with windows lacking features, which could subsequently lead to difficulties in training the upper-layer parameters. The term $h$ is introduced to counteract the effects of intensity scaling, which will be discussed further in the following text. In short, we hope that the Window Extractor can fully capture various features from the high dynamic range data of CT scans without missing important areas at the very beginning of training.

\textbf{Second,} we use $\beta = (\textit{tanh}(g)+1) / h$, where $h$ is the aforementioned range rectification coefficient, $g$ is the dynamic range coefficient. $h$ limits the value domain that one window can cover and ensures that one single window will not overly extract too much features, which will downgrade the window to a whole-window Tanh activation. When it is set to a lower value, the output will have a more intense derivative. According to the discussion in \cref{sec:WinE_OriTanh}, this is equal to a narrower window. $h$ is defaults to $\mathfrak{R} / N$, where $N$ is the number of the paralleled window extractor and $\mathfrak{R}$ is source HU value range of CT. In terms of $g$, it also acts as a coefficient to control window width, just like $h$.

Noted that the numerator of $\beta$ is directly proportional to the mapping response. When the numerator is very close to zero, the response will be very weak, leading to the vanishing gradient problem. At the other extreme, when the numerator is too large, the window becomes too narrow, resulting in a higher numerical variance of the extracted image, and a small shift in the window location can lead to dramatically different features being extracted. In this case, the upper neural network will become very aggressive during learning. Imagine this scenario: "In two consecutive optimizations iteration, two very different strong signals have the same label," which can lead to gradients that are more prone to explode. In this case, we warp the learnable $g$ with $\textit{tanh}(g)+1$. This is ensures 1) the gradient will not vanish at this step as the numerator will always be positive, and 2) $g$ needs to be much larger to gain a narrow window. The model needs consistent judgment across multiple samples, which indicating that there do exist a strong feature that may be helpful to the downstream task.

The mathematical process is outlined in \cref{eq:tauS}.
\begin{equation}
    \label{eq:tauS}
    \tau(\mathcal{S}) = \frac{\textit{tanh}(g)+1}{h} \cdot (\mathcal{S}-m+hd)
\end{equation}

\subsubsection{Integration into Tanh}

Our design approach is to create a pre-bias mapping for the Tanh function, so replacing $\textit{tanh}(x)$ with $\textit{tanh}(\tau(x))$ would suffice. We further add a global response offset $k$, which acts like a bias of a linear layer, and can globally adjust the weight of the window across all parallel windows. $a$ and $b$ correspond to the fine-tuning in the negative and positive directions, respectively, their value are limited to , and will have a decreased control intensity as they increase. Integrating the above discussion, the modified Tanh function can be expressed as \cref{eq:window_extractor}.

\begin{equation}
    \label{eq:window_extractor}
    \mathcal{W}(\mathcal{S}) = \frac{(a+1)e^{\tau(\mathcal{S})} - (b+1)e^{-\tau(\mathcal{S})}}{(a+1)e^{\tau(\mathcal{S})} + (b+1)e^{-\tau(\mathcal{S})}} + k
\end{equation}

\subsubsection{Mathematical Property Analysis}

As $\mathcal{W}$ is the first step in the entire forward process, any numerical unstability will be amplified at each higher neural layer, so it is necessary to pay great attention to the stability of the mapping process to ensure that different upper neural networks can be trained efficiently.

$a, b, d$ control the window location. $a, b$ have nonlinearly attenuated control intensity (\cref{eq:WinExt_S0_partial_a}, \cref{eq:WinExt_S0_partial_b}), which is considered as small scale control. Meanwhile, $d$ controls the window linearly (\cref{eq:tauS}), which is considered as large scale control. Thus, we can introduce two different scales of window level control for the projection and no divergence points exists.
\begin{equation}
    \label{eq:WinExt_S0}
    \theta = \mathcal{W}(0) = \frac{a-b}{a+b+2} + k
\end{equation}
\begin{equation}
    \label{eq:WinExt_S0_partial_a}
    \frac{\partial \theta}{\partial a} {\bigg |}_{a \to +\infty} = \frac{2b+2}{(a+b+2)^2} {\bigg |}_{a \to +\infty} = 0
\end{equation}
\begin{equation}
    \label{eq:WinExt_S0_partial_b}
    \frac{\partial \theta}{\partial b} {\bigg |}_{b \to +\infty} = \frac{-2a-2}{(a+b+2)^2} {\bigg |}_{b \to +\infty} = 0
\end{equation}

The response feature of Window Extractor is still similar to \textit{Tanh}. To confirm this point, here we conducted a mathematical analysis of the function’s characteristics.

1) First order monotonic increasing, \cref{eq:WinE_resp_feat} is always positive.
\begin{equation}
    \frac{d\mathcal{\tau}}{d\mathcal{S}} = \frac{\textit{tanh}(g)+1}{h} > 0
\end{equation}
\begin{equation}
    \label{eq:WinE_resp_feat}
    \frac{d\mathcal{W}}{d\mathcal{S}} = \frac{2\tau^\prime(\mathcal{S})(b+1)e^{-\tau(\mathcal{S})}}{[(a+1)e^{\tau(\mathcal{S})}+(b+1)e^{-\tau(\mathcal{S})}]^2} > 0
\end{equation}

2) Similar sign trait in the second derivative. The sign of \cref{eq:WinE_sec_derv} is determined by the last term \cref{eq:WinE_sec_derv_sign} in the numerator. The derivative of it is always negtive \cref{eq:WinE_sec_derv_sign_derv}. When further consider \cref{eq:WinE_sec_derv_sign_zero}, we can conclude that the $\frac{d^2\mathcal{W}}{d\mathcal{S}^2}$ has only one root, and $\mathcal{W}(\mathcal{S})$ is concave to the left and convex to the right at the root.
\begin{equation}
    \label{eq:WinE_sec_derv}
    \frac{d^2\mathcal{W}}{d\mathcal{S}^2} = \frac{2(b+1) e^{-\tau(\mathcal{S})} \cdot \tau^{\prime2}(\mathcal{S}) \cdot \mathit{f}(\mathcal{S})}
    {((a+1)e^{\tau(\mathcal{S})}+(b+1)e^{-\tau(\mathcal{S})})^3}
\end{equation}
\begin{equation}
    \label{eq:WinE_sec_derv_sign}
    \mathit{f}(\mathcal{S}) = 1-\tau(\mathcal{S})-2(a+1)e^{\tau(\mathcal{S})}+2(b+1)e^{-\tau(\mathcal{S})}
\end{equation}
\begin{equation}
    \label{eq:WinE_sec_derv_sign_zero}
    \mathit{f}(-\infty) \cdot \mathit{f}(+\infty) < 0 \leftrightarrow \exists \lambda, \mathit{f}(\lambda) = 0
\end{equation}
\begin{equation}
    \begin{split}
        \label{eq:WinE_sec_derv_sign_derv}
        \frac{d\mathit{f}}{d\mathcal{S}} &= -\tau^\prime(\mathcal{S}) - \tau^\prime(\mathcal{S}) \cdot 2(a+1)e^{\tau(\mathcal{S})} - 2(b+1)e^{-\tau(\mathcal{S})} \cdot \tau^\prime(\mathcal{S})\\
        &= -\tau^\prime(\mathcal{S}) \cdot (1 + 2(a+1)e^{\tau(\mathcal{S})} + 2(b+1)e^{-\tau(\mathcal{S})}) < 0
    \end{split}
\end{equation}

3) Similar Range. The range of the \textit{Tanh} function is $(-1, 1)$ while the Window Extractor's being $(k-1, k+1)$.

\subsection{Tanh-Based Post Rectificator}
\label{sec:Method_TRec}

Upon synthesizing the aforementioned analysis, it is concluded that $\mathcal{W}$ exhibits a deficiency in non-linear degrees of freedom. This implies a limitation in its capacity to concurrently attend to a spectrum of signal intensities, being primarily focused on the enhancement of values in the proximity of $\lambda$. 

At this juncture, the adoption of an auxiliary \textit{Tanh} response in conjunction with residual connections is instrumental in augmenting the signals emanating from each Window Extractor. We introduce 1) Multiple interested location stored in $\mathcal{R} \in \mathbb{R}^\kappa$ and 2) Focus intensity of each location stored in $\mathcal{K} \in \mathbb{R}^\kappa$. The mathematical process is outlined as follows:
\begin{equation}
    \mathcal{A}(\mathcal{S}) = \mathcal{K} \cdot \textit{tanh}(\mathcal{R} + \mathcal{I}_\kappa \otimes \mathcal{W}(\mathcal{S}))
\end{equation}
\begin{equation}
    \mathcal{O}_{\textit{TRec}}(\mathcal{S}) = \mathcal{W}(\mathcal{S}) + \mathcal{A}(\mathcal{S})
\end{equation}
where $\mathcal{I}_\kappa$ is the identity matrix of size $\kappa$.

Similar to \cref{eq:tauS}, $\mathcal{R}$ introduces location offsets for each tanh rectification member. $\mathcal{I}_\kappa \otimes \mathcal{W}(\mathcal{S})$ is designed for more efficient calculation, the original response will be copied $\kappa$. Then, the location offset and intensity can be calculated in parallel. Intuitively, the response curve after rectification will behave as if several tanh sub-shapes are superimposed on a typical tanh shape.

\subsection{Paralleled Windows and Fusion}
\label{sec:Method_CrsF}

By using one $\mathcal{W}$ together with one $\mathcal{A}$, we get one enhanced HU signal channel $\mathcal{O}_{\textit{TRec}}$. As has been mentioned above, single $\mathcal{O}_{\textit{TRec}}$ only obtain a narrow value range. So we use multiple paralleled signal stream with the same enhance method to generate multi-channel signal $\mathcal{O}_{\textit{TRec1}}, \mathcal{O}_{\textit{TRec2}}, \ldots, \mathcal{O}_{\textit{TRecN}}$.

The generation of distinct channels from a single signal source echoes the application of the well-established self-attention technique, which dynamically determines the relevance of channels and enables data interchange based on this assessment. However, the inherent absence of learnable parameters in self-attention precludes the determination of a constant weight matrix for channel-wise data transfer. Considering the strict definition and implicit connotations of HU in CT imaging, it is imperative to avoid introducing steps with limited interpretability during channel-wise data exchange. Consequently, we explicitly define a learnable weight matrix $\mathbf{H} \in \mathbb{R}^{N \times N}$ and using the following mathematical process:
\begin{equation}
    \mathcal{O}_{\textit{TRec}} = \textit{softmax}(\mathbf{H})  \cdot 
    \begin{bmatrix}
        \mathcal{O}_{\textit{TRec1}} \\
        \mathcal{O}_{\textit{TRec2}} \\
        \vdots \\
        \mathcal{O}_{\textit{TRecN}}
    \end{bmatrix}
\end{equation}
noted that the softmax operation is applied to each row of the matrix $\mathbf{H}$.

After the fusion operation, we concatenate all signal channels to form the final output $\mathbf{P}^{(N)}(\mathcal{S}\in \mathbb{R}^{C \times Z \times Y \times X}) \in \mathbb{R}^{(N \times C) \times Z \times Y \times X}$, where $C$ is the channel number of the original CT image and is usually 1.

Now, without any human intervention, the existing neural network backbone can automatically perceive fine signal changes from different HU domains in parallel channels and backpropagate gradients to the window extractor, automatically determining the optimal signal range for the target of interest.

\section{Experiments and Results}
\label{sec:Exp}

\subsection{Datasets and Settings}
\label{sec:Exp_Dataset}

In this study, we primarily utilize several widely recognized public datasets, including Totalsegmentator \citep{dantonoli2024totalsegmentator}, FLARE 2023 \citep{FLARE2023}, CT-ORG \citep{Rister2020}, AbdomenCT-1K \citep{MaAbdomenCT1K}, KiTS23 \citep{heller2023kits19, heller2023kits21}, and ImageTBAD \citep{YaoImageTBAD}.
Then, we introduce a proprietary CT imaging dataset to validate the sensitivity of the proposed method. Our private data is not included during training, so can be used to observe our method's generalization performance. The IRB approval is available if required.

\begin{table}
    \raggedright
    \caption{All datasets used in this study, including 6 public datasets and 1 proprietary dataset. Their tasks covers universal segmentation, organ segmentation and disease segmentation.}
    \label{tab:datasets}

    \begin{tabular}{llll}
        \hline
        Dataset & Used Series & Used Classes & Task Description\\
        \hline
        Totalsegmentator & 1228 & 119 & Organ Segmentation\\
        CT-ORG & 140 & 6 & Organ Segmentation\\
        AbdomenCT-1K & 1000 & 5 & Organ Segmentation\\
        FLARE 2023 & 2199 & 15 & Organ and General Tumor \\
        KiTS23 & 489 & 4 & Kidney Tumor Segmentation\\
        ImageTBAD & 100 & 4 & Type-B Aortic Dissection\\
        Private & 700 & 2 & Gastric Cancer\\
        \hline
    \end{tabular}
\end{table}

In our experiments, we majorlly use MedNeXt \citep{MedNeXt2023} proposed by nnUNet \citep{RoynnUNet2023} project members. The reason for choosing this model is its high reproducibility, which has been integrated into the nnUNet framework, a benchmark training pipeline widely used in the medical field. Its structure maintains the classic Encoder-Decoder form, focusing mainly on convolutional extraction, with almost no complex design. This allows it to achieve acceptable performance on different tasks. We hope our method can achieve as high a level of universality as possible, so using a typical model as an experimental subject is a good choice.

The MedNeXt is set to 3D mode for patch-based volume segmentation task, and the patch size is $64 \times 64 \times 64$. We use MMEngine \citep{mmengine2022} framework to construct all experiments in section, the detailed configurations and data-preprocessing methods are included in our github repo.

\subsection{Large-Scale Multi-Target Organ Segmentation}
\label{sec:Exp_LargeScale}

Large datasets have a significantly greater number of samples and may contain more annotated instances and categories, thus involving features of different HU subdomains. We allege that using large datasets can more significantly verify whether our method can effectively perform sub-window positioning and extraction.

\subsubsection{Totalsegmentator}

We train on the Totalsegmentator dataset \citep{dantonoli2024totalsegmentator}, which aims to identify 119 human tissues or structures in whole-body CT scans. These classes span the lungs (low HU region), digestive organs or tract (near-zero HU region), bone tissue (high HU region), etc., which can well measure whether our proposed method helps neural networks adaptively extract information from different sub-windows. The voxel spacing is aligned to $2 \times 2 \times 2 (mm)$.

We trained the MedNeXt model with 500K iterations, 4 $\times$ RTX 4090 GPU with batchsize 8 on each of them. The best checkpoint during training is used for final testing. For large-scale multi-classification tasks, a consensus on a specific window setting has not yet been reached. Consequently, we adopt instance normalization as our baseline. Our training process maintains all other preprocessing steps and neural network configurations unchanged. 

The results are shown in \cref{tab:TotalsegmentatorTrain}. It indicates that in the absence of the Auto Window feature, the Dice achieved 78+, with the Recall exhibiting a superiority over Precision. This pattern suggests that the model is inclined towards over-segmentation. After deploying the proposed Auto Window, we see a boost in accuracy across the board. While Recall remains marginally superior to Precision, the two are closer, indicating a greater robustness in the current predictions. The utilization of 8 Auto Windows has led to enhancements in certain metrics when compared to 4 Auto Windows, although the overall impact remains modest.
\begin{table}[htbp]
    \raggedright
    \caption{Performance comparison of different configurations on the Totalsegmentator dataset. The proposed method achieves the better accuracy. The gap between Recall and Precision narrows after the Auto Window is deployed. The overall performance between 4 and 8 windows is similar.} 
    \label{tab:TotalsegmentatorTrain}

    \begin{tabular}{lllll}
        \hline
        Method & Dice & IoU & Recall & Precision\\
        \hline
        Instance Norm  & 78.41 & 67.20 & 85.66 & 73.59\\
        4 Auto Windows & \textbf{90.66} & 84.78 & 91.94 & \textbf{90.40}\\
        8 Auto Windows & 90.47 & \textbf{85.01} & \textbf{92.57} & 90.03\\
        \hline
    \end{tabular}
\end{table}

Upon analyzing the Class-Wise performance in \cref{tab:TotalsegmentatorClassWise}, it is evident that the neural network encountered challenges in identifying the adrenal gland without the implementation of Auto Window, yielding a Dice score of approximately $40$. However, the incorporation of Auto Window lead to a remarkable enhancement in recognition performance for this category, with a Dice score exceeding $87$. Similar improvements are noted in the identification of the common carotid artery, iliac artery, gallbladder, duodenum, etc.. The most notable enhancement is in Precision, indicating a substantial reduction in the likelihood of the model incorrectly classifying other regions as the target.
\begin{table}[htbp]
    \raggedright
    \caption{Class-Wise performance of those classes that were originally difficult to predict in Totalsegmentator Dataset. Auto Window has rendered the model’s predictions for these classes essentially viable. Gain is determined based on the superior configuration among the two Window options. } 
    \label{tab:TotalsegmentatorClassWise}

    \begin{tabular}{llllll}
        \hline
        Class & Metric & Instance Norm & 4 Auto Windows & 8 Auto Windows & Gain\\
        \hline
        \multirow{3}{*}{Adrenal Gland} 
        & Dice & 38.33 & 87.25 & 85.67 & \textbf{+127\%}\\
        & Recall & 67.51 & 89.45 & 87.60 & \textbf{+32\%}\\
        & Precision & 27.00 & 85.15 & 83.84 & \textbf{+215\%}\\
        \hline
        \multirow{3}{*}{Iliac Artery} 
        & Dice & 53.89 & 86.66 & 88.20 & \textbf{+64\%}\\
        & Recall & 75.41 & 91.79 & 92.80 & \textbf{+23\%}\\
        & Precision & 42.04 & 82.10 & 84.05 & \textbf{+100\%}\\
        \hline
        \multirow{3}{*}{Gallbladder} 
        & Dice & 51.86 & 89.37 & 89.47 & \textbf{+73\%}\\
        & Recall & 73.43 & 95.53 & 94.98 & \textbf{+30\%}\\
        & Precision & 40.09 & 83.95 & 84.56 & \textbf{+111\%}\\
        \hline
        \multirow{3}{*}{Duodenum} 
        & Dice & 58.20 & 89.32 & 89.63 & \textbf{+54\%}\\
        & Recall & 79.19 & 89.49 & 90.48 & \textbf{+14\%}\\
        & Precision & 46.00 & 89.15 & 88.79 & \textbf{+94\%}\\
        \hline
        \multirow{3}{*}{Pancreas} 
        & Dice & 57.47 & 92.44 & 92.78 & \textbf{+61\%}\\
        & Recall & 73.18 & 93.11 & 94.42 & \textbf{+29\%}\\
        & Precision & 47.32 & 91.79 & 91.20 & \textbf{+94\%}\\
        \hline
    \end{tabular}
\end{table}

\begin{figure}[tbp]
    \centering
    \begin{subfigure}[b]{0.45\linewidth}
        \includegraphics[width=\linewidth]{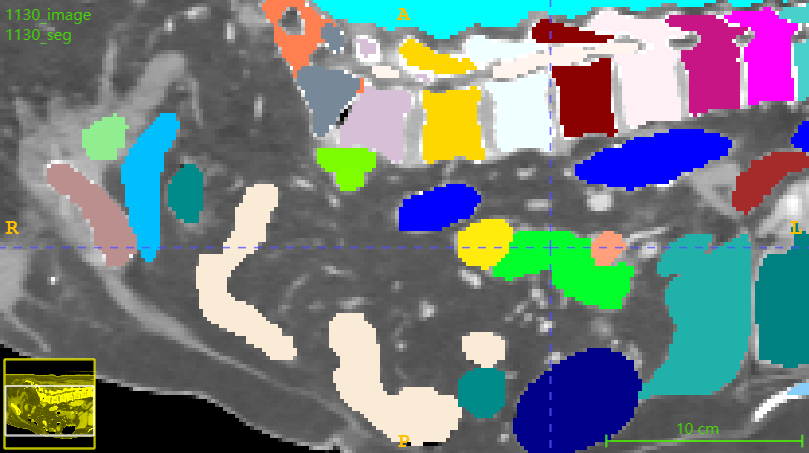}
        \caption{Ground Truth Sagittal Whole Window}
        \label{fig:HardInfer_Totalseg_Sagittal_Image}
    \end{subfigure}
    \begin{subfigure}[b]{0.45\linewidth}
        \includegraphics[width=\linewidth]{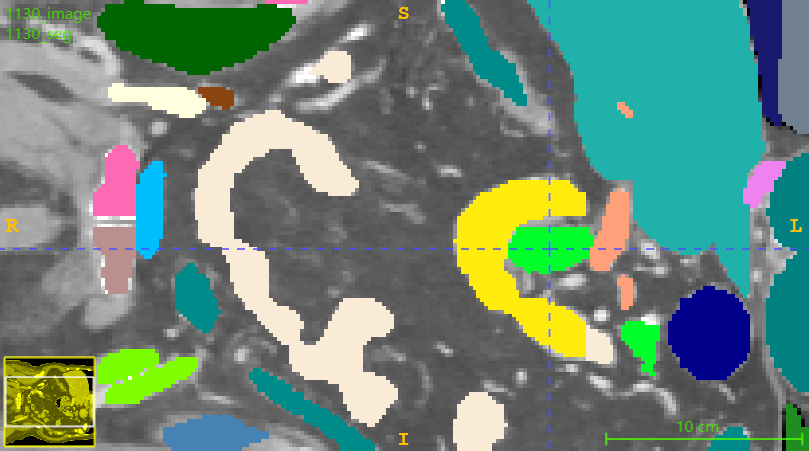}
        \caption{Ground Truth Coronoal}
        \label{fig:HardInfer_Totalseg_Coronoal_Image}
    \end{subfigure}

    \begin{subfigure}[b]{0.45\linewidth}
        \includegraphics[width=\linewidth]{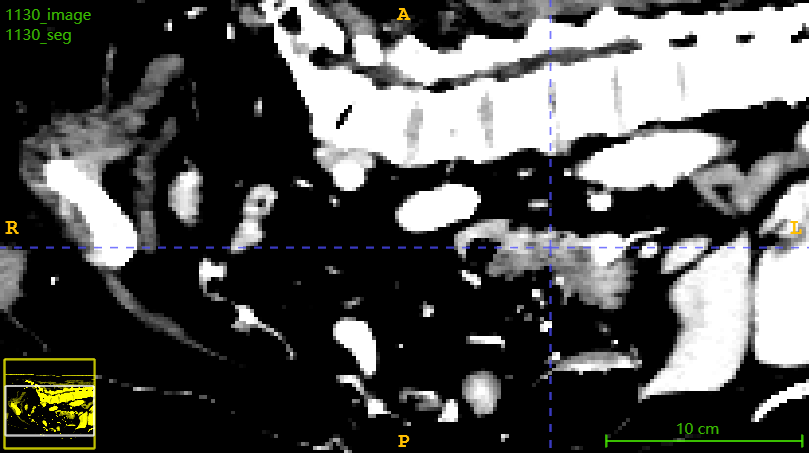}
        \caption{Auto Window Sagittal}
        \label{fig:HardInfer_Totalseg_Sagittal_AutoWindow}
    \end{subfigure}
    \begin{subfigure}[b]{0.45\linewidth}
        \includegraphics[width=\linewidth]{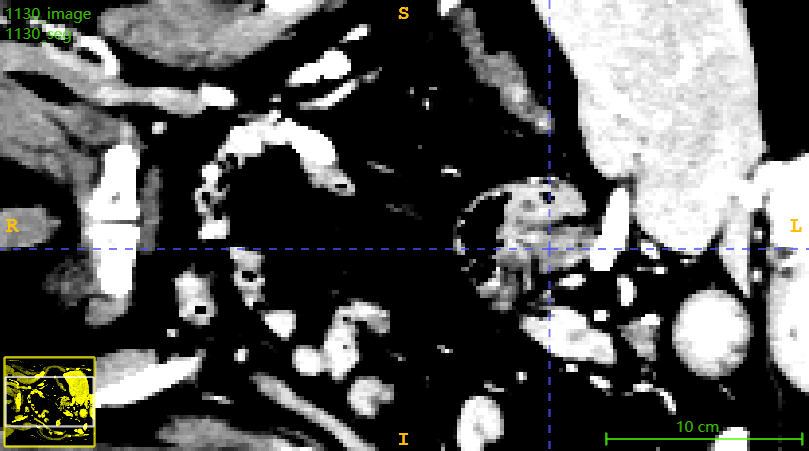}
        \caption{Auto Window Coronoal}
        \label{fig:HardInfer_Totalseg_Coronoal_AutoWindow}
    \end{subfigure}

    \begin{subfigure}[b]{0.45\linewidth}
        \includegraphics[width=\linewidth]{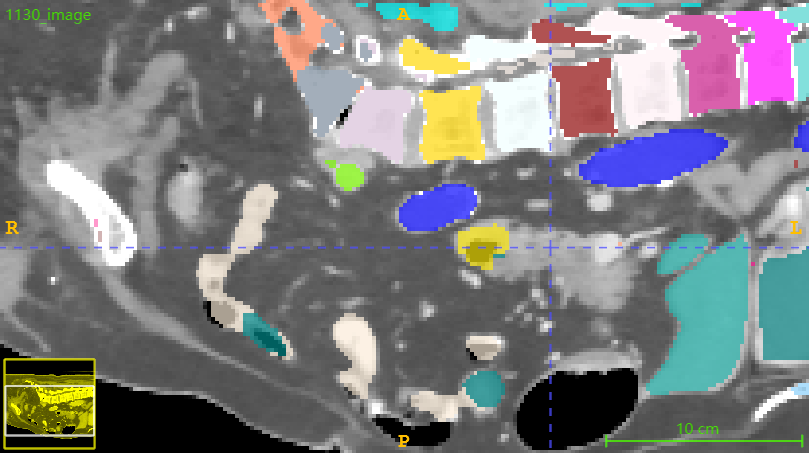}
        \caption{Vanilla Prediction Sagittal}
        \label{fig:HardInfer_Totalseg_Sagittal_Pred_InstanceNorm}
    \end{subfigure}
    \begin{subfigure}[b]{0.45\linewidth}
        \includegraphics[width=\linewidth]{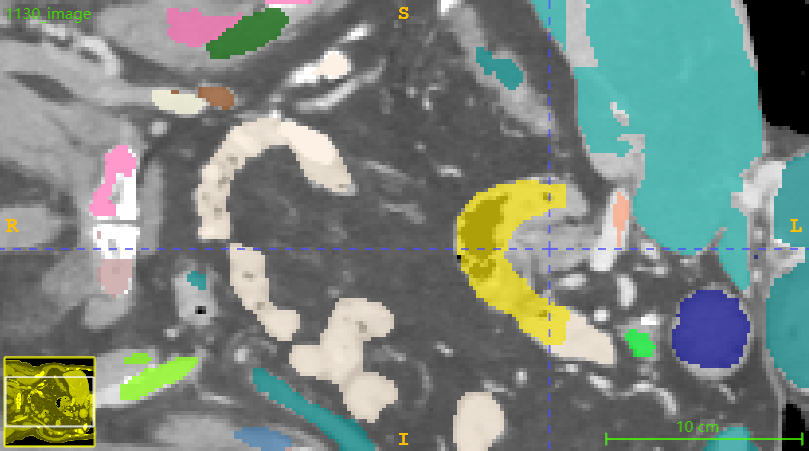}
        \caption{Vanilla Prediction Coronoal}
        \label{fig:HardInfer_Totalseg_Coronoal_Pred_InstanceNorm}
    \end{subfigure}

    \begin{subfigure}[b]{0.45\linewidth}
        \includegraphics[width=\linewidth]{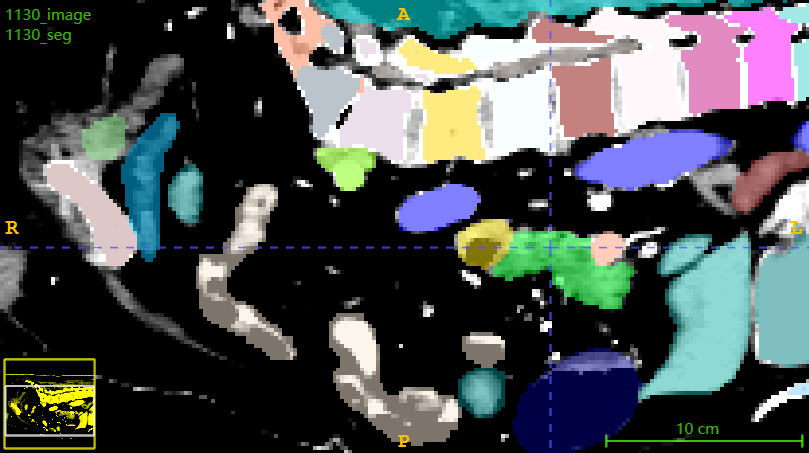}
        \caption{Auto Window Prediction Sagittal}
        \label{fig:HardInfer_Totalseg_Sagittal_Pred_AutoWindow}
    \end{subfigure}
    \begin{subfigure}[b]{0.45\linewidth}
        \includegraphics[width=\linewidth]{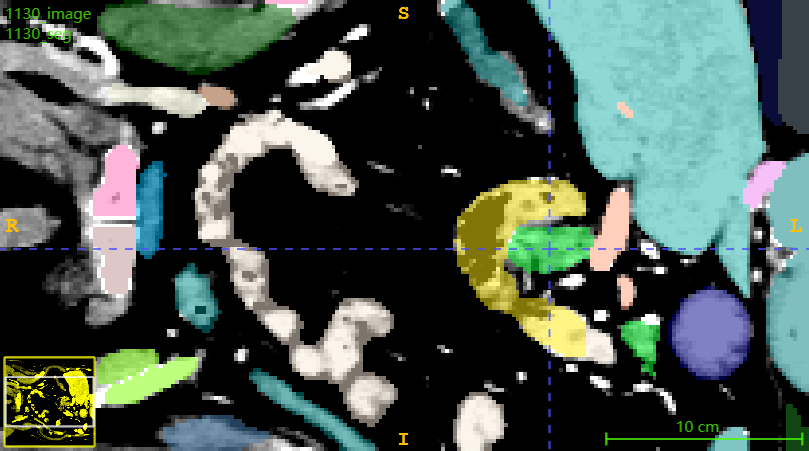}
        \caption{Auto Window Prediction Coronoal}
        \label{fig:HardInfer_Totalseg_Coronoal_Pred_AutoWindow}
    \end{subfigure}

    \caption{The original CT image and the corresponding segmentation results of the Duodenum and Iliac Artery. The neural network has difficulty distinguishing between these two structures when using whole-window. The duodenum annotation/prediction is represented in a luminous yellow, while the pancreas is in a luminous green. In the given sample, the model lacking the Auto Window fails to accurately segment the pancreas. It's incapable of generating usable predictions, resulting in the pancreas area being misclassified as background.}
    \label{fig:HardInfer_Totalseg}
\end{figure}

We conducted an analysis to understand the substantial enhancement in the duodenum and pancreas categories as an example in \cref{fig:HardInfer_Totalseg}. Anatomically, the proximity of the Duodenum and Pancreas, coupled with their minimal distinction in HU values, poses a challenge in easily demarcating the boundary between these two structures (\cref{fig:HardInfer_Totalseg_Sagittal_Image}), even for experienced physicians interpreting medical images. When using whole-window for feature extraction, the HU gap between them will be too close to be embeded differently. In this scenario, the neural network is forced to tell a difference between two similar feature vectors, thus causing learning difficulties. The Auto Window, however, can help amplify the subtle boundary feature, and provide more easy-to-learn knowledge for the neural network (\cref{fig:HardInfer_Totalseg_Sagittal_AutoWindow}). Meanwhile, the Auto Window will filter irrelated informations, so the channels from this Auto Window to Neural Embedding Layer will be more focused on the key target.

\begin{figure}[tbp]
    \centering
    \begin{subfigure}[b]{0.45\linewidth}
        \includegraphics[width=\linewidth]{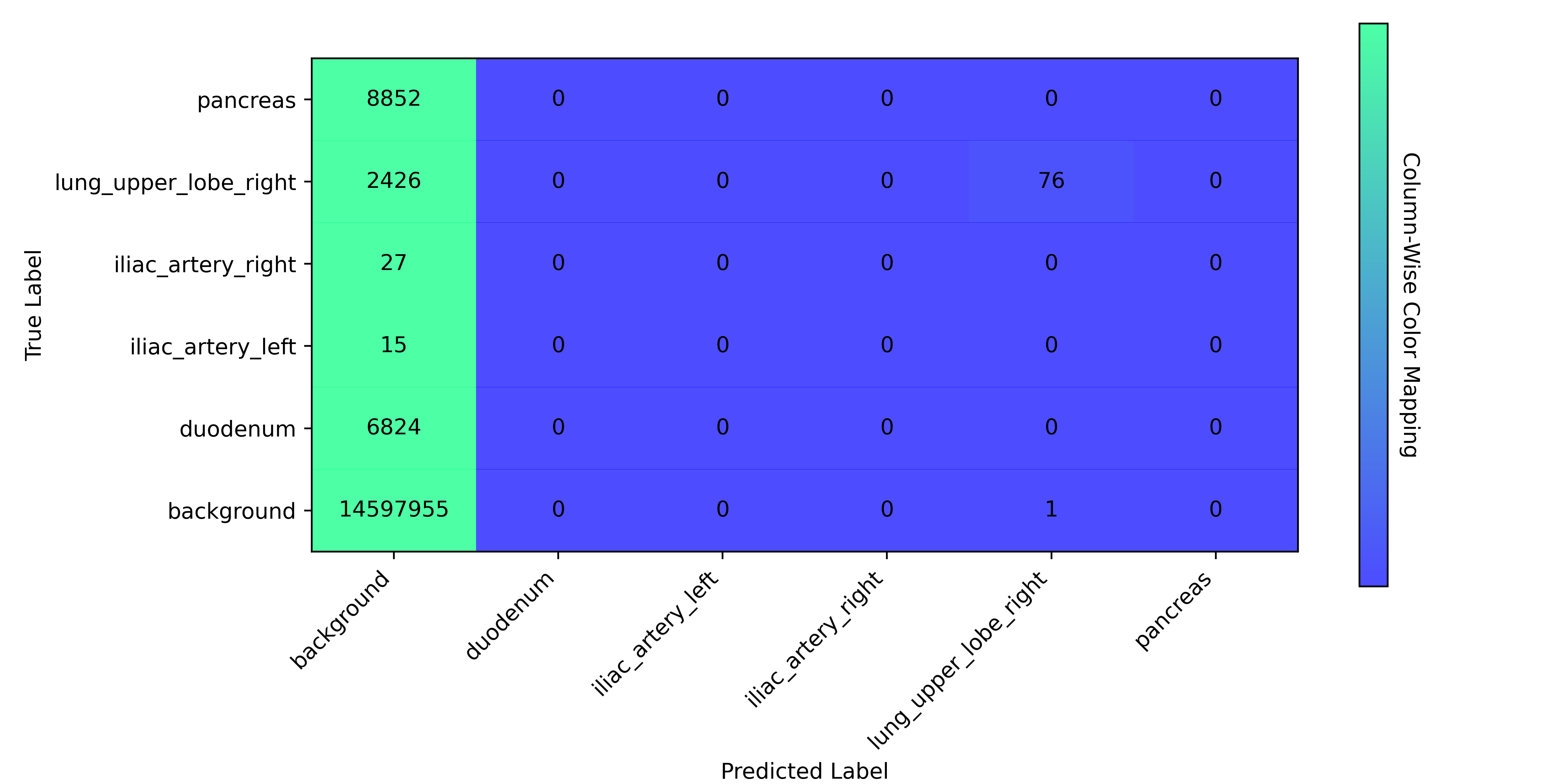}
        \caption{Instance Norm}
        \label{fig:HardInfer_Totalseg_Confusion_InstanceNorm}
    \end{subfigure}
    \begin{subfigure}[b]{0.45\linewidth}
        \includegraphics[width=\linewidth]{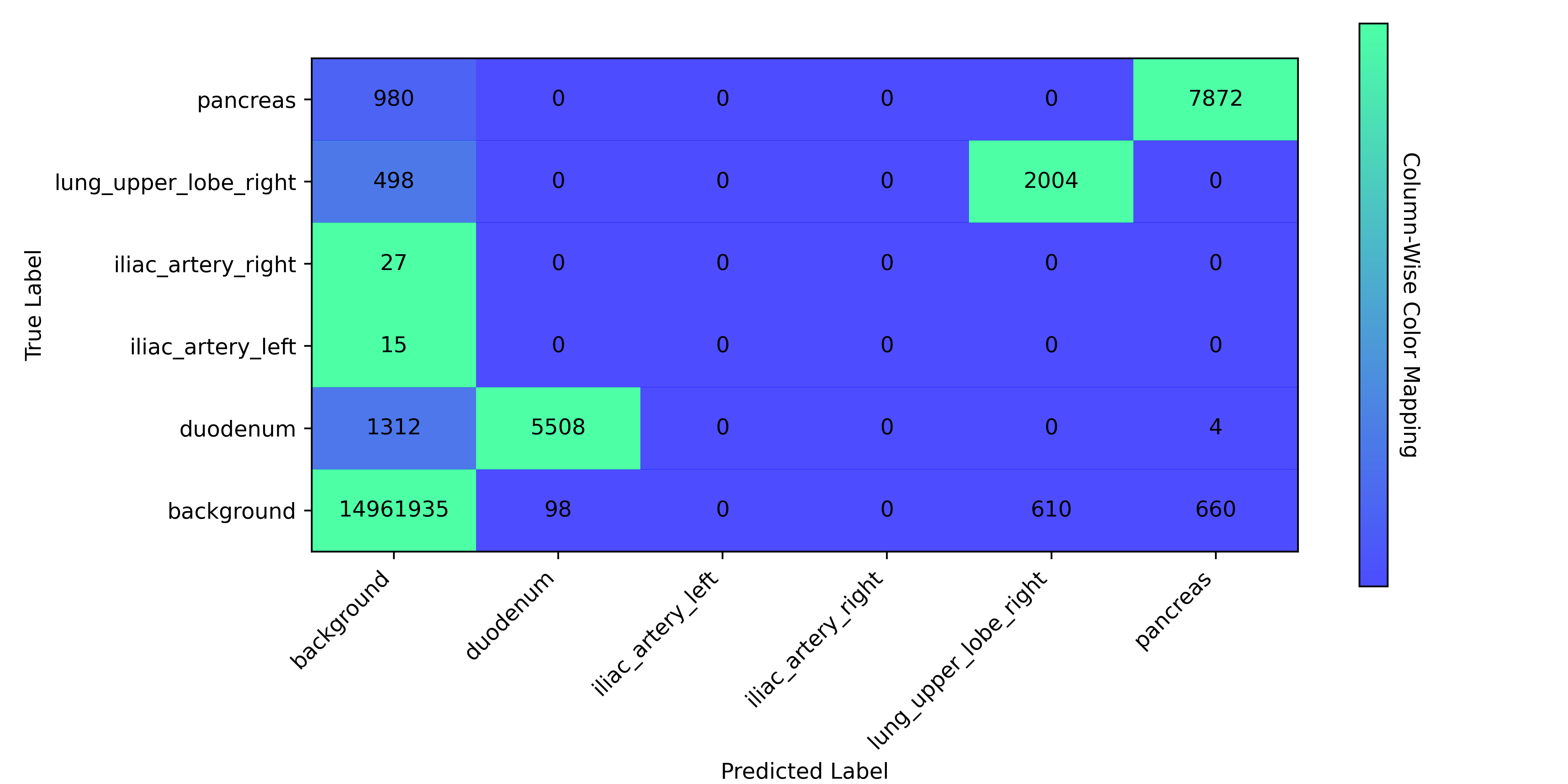}
        \caption{8 Auto Windows}
        \label{fig:HardInfer_Totalseg_Confusion_8AutoWindow}
    \end{subfigure}
    \caption{Confusion matrix of the hard classes in the Totalsegmentator dataset. The Auto Window enables the neural network to detecting them, rather than just failing. Nevertheless, certain classes exhibit insensitivity to them.}
    \label{fig:HardInfer_Totalseg_Confusion}
\end{figure}

Utilizing the capabilities of Auto Window, the pancreas and duodenum can be effectively differentiated with high precision, yielding Dice coefficients of $92$ and $89$ respectively (\cref{tab:TotalsegmentatorClassWise}). This represents an improvement over previous methods, which often struggled to distinguish between these two structures, resulting in incomplete recognition in certain samples (\cref{fig:HardInfer_Totalseg_Sagittal_Pred_InstanceNorm}). Previously, effective execution of this recognition task necessitated the deployment of a specialized neural network for an individual analytical process, which in turn, amplified both the design complexity as well as the computational overhead. An analysis of the prediction confusion (\cref{fig:HardInfer_Totalseg_Confusion}) also reveals enhancements in multiple classes that were originally challenging to segment. The neural network is less likely to classify them as background or to misclassify them as other closely related classes, also indicating that valuable knowledge has been learned from the Auto Window. Nevertheless, there are persistent challenges with certain classes that remain difficult to predict even when employing the Auto Windows. This could be attributed to the insufficient quantity of effective annotations for these classes, resulting in a significant imbalance that may cause the neural network to disregard them during training.

\subsubsection{FLARE 2023}

Although Totalsegmentor dataset has over 100 targets, it does not include any disease targets. So we have incorporated the FLARE 2023 dataset as a supplementary resource. This dataset is primarily aimed at the identification and segmentation of abdominal organs and their cancerous tissues, comprising 13 distinct organ categories and a general cancer target. Training configurations remain the same as those used for the Totalsegmentor dataset. The results are shown in \cref{tab:FLARE2023Train}.
\begin{table}[htbp]
    \raggedright
    \caption{Performance comparison of different configurations on the FLARE 2023 dataset.} 
    \label{tab:FLARE2023Train}

    \begin{tabular}{lllll}
        \hline
        Method & Dice & IoU & Precision & Recall\\
        \hline
        Instance Norm  & 29.14 & 21.27 & 64.09 & 24.82\\
        4 Auto Windows & \textbf{76.12} & \textbf{62.86} & \textbf{88.76} & \textbf{68.13}\\
        8 Auto Windows & 69.67 & 55.57 & 81.22 & 63.25\\
        \hline
    \end{tabular}
\end{table}

\subsection{Small-Scale Local Analysis}
\label{sec:Exp_SmallScale}

\subsubsection{Public Dataset}

For most diagnostic models for specific diseases, a single window is often used for analysis. In scenarios where multiple windows are improperly configured, there is a risk of incorporating superfluous information, which can adversely affect the training process of the model. Consequently, it is imperative to evaluate the robustness of the Auto Window in small tasks. Under optimal configuration, the performance with Auto Window should be comparable to, if not superior to, that achieved with windows determined through empirical methods.

We utilized two regional organ datasets and three disease identification datasets to test the analytical performance of our method for specific categories. These datasets contain six or fewer categories, which is much fewer compared to the large datasets used in \cref{sec:Exp_LargeScale}. These datasets encompass the identification of blood vessels, kidneys, and certain abdominal organs, so we use mediastinal window $[40 \pm 200]$ to examine the tradition window setting method as a baseline. For KiTS23 dataset, we fine-tune the window to $[20 \pm 200]$. This manual window is widely used in clinical practice and can observe the abdomen targets. The experiments conducted on the Small-Scale dataset are executed utilizing an RTX 4070 Ti and an RTX 2070 Super.

\begin{figure}[tbp]
    \centering
    \begin{subfigure}[b]{0.32\linewidth}
        \includegraphics[width=\linewidth]{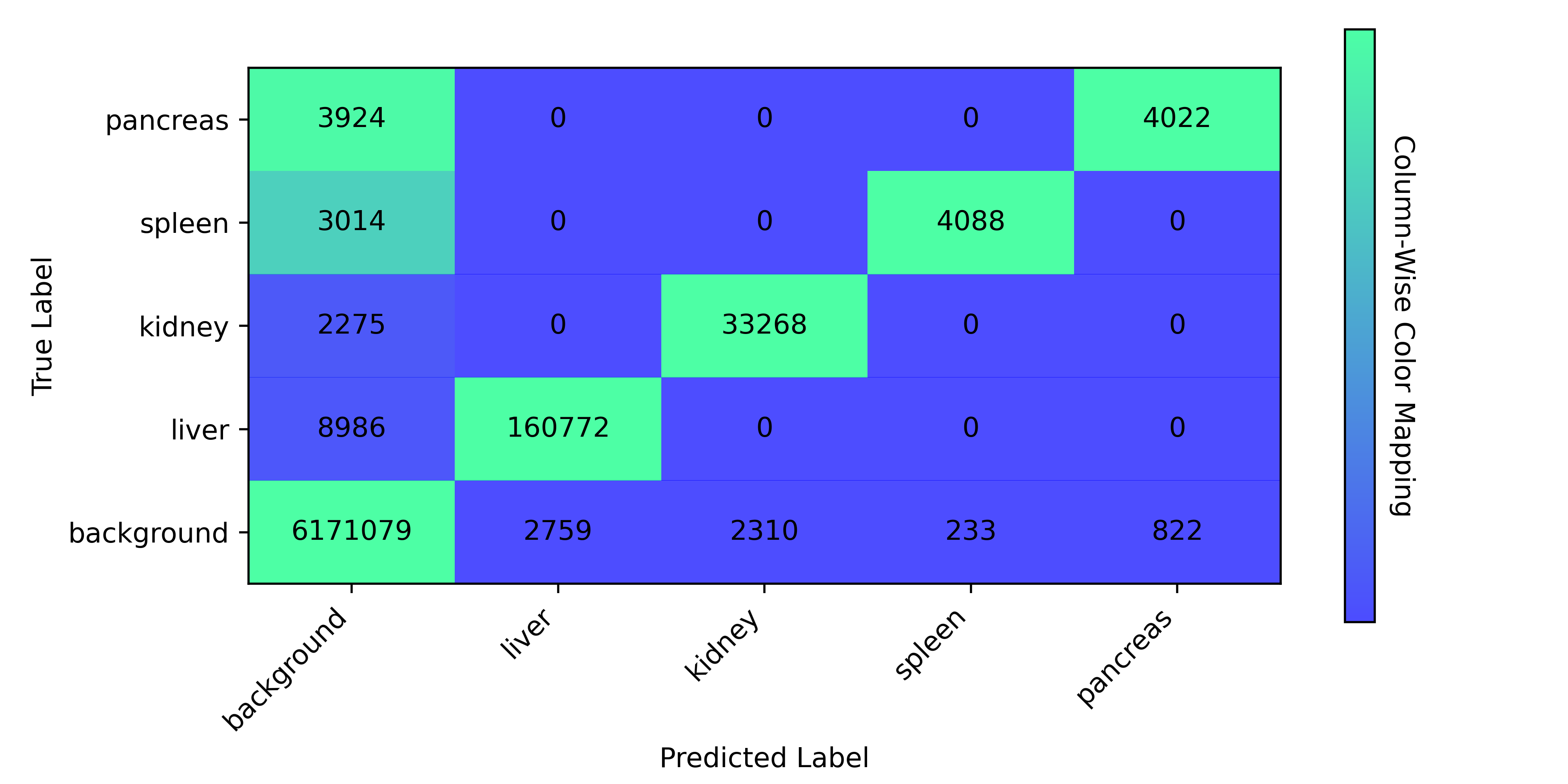}
        \caption{Instance Norm}
        \label{fig:SmallScaleExp_cm_AbdomenCT1K_InstNorm}
    \end{subfigure}
    \begin{subfigure}[b]{0.32\linewidth}
        \includegraphics[width=\linewidth]{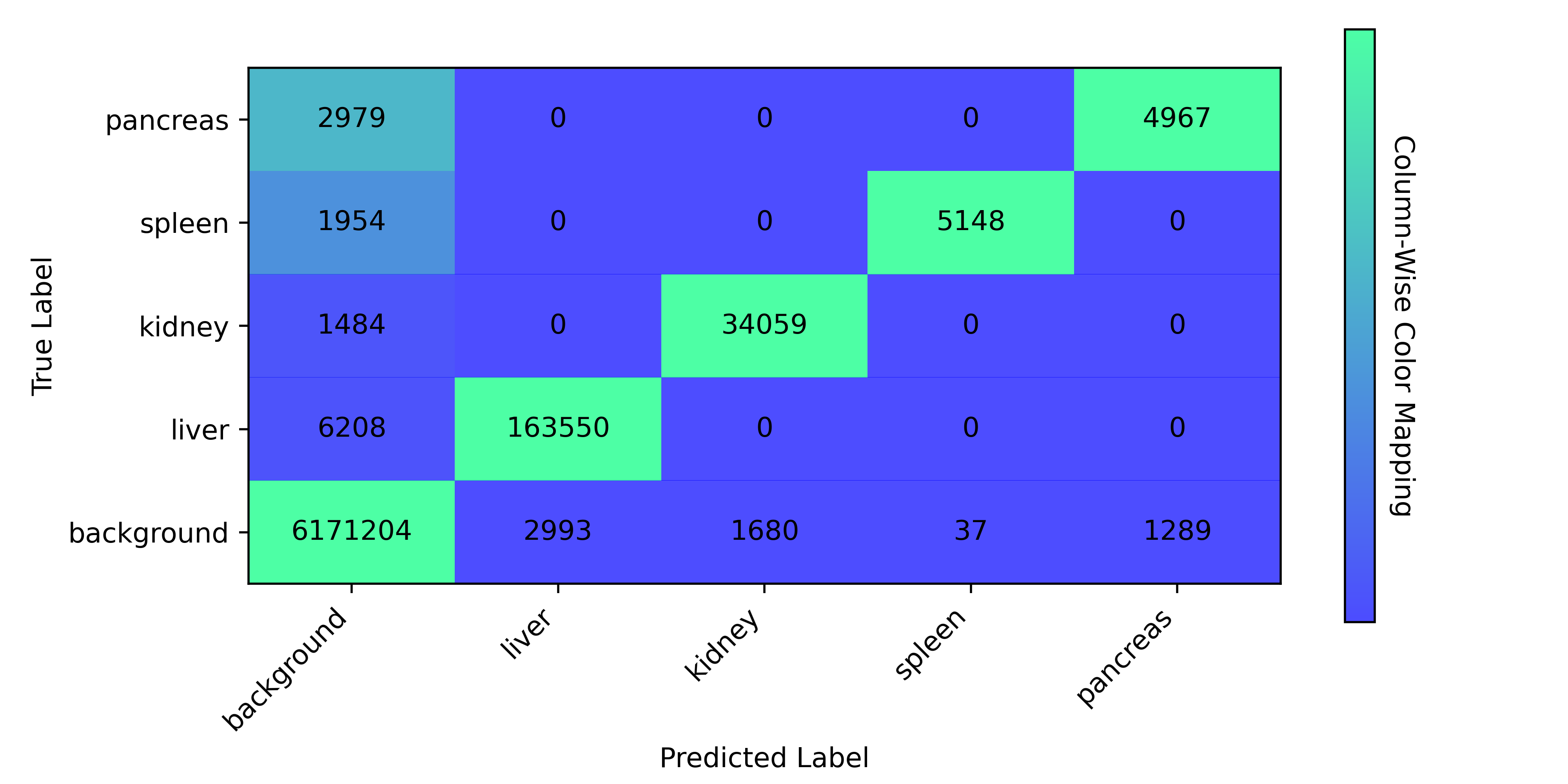}
        \caption{Manual Window}
        \label{fig:SmallScaleExp_cm_AbdomenCT1K_ManualWin}
    \end{subfigure}
    \begin{subfigure}[b]{0.32\linewidth}
        \includegraphics[width=\linewidth]{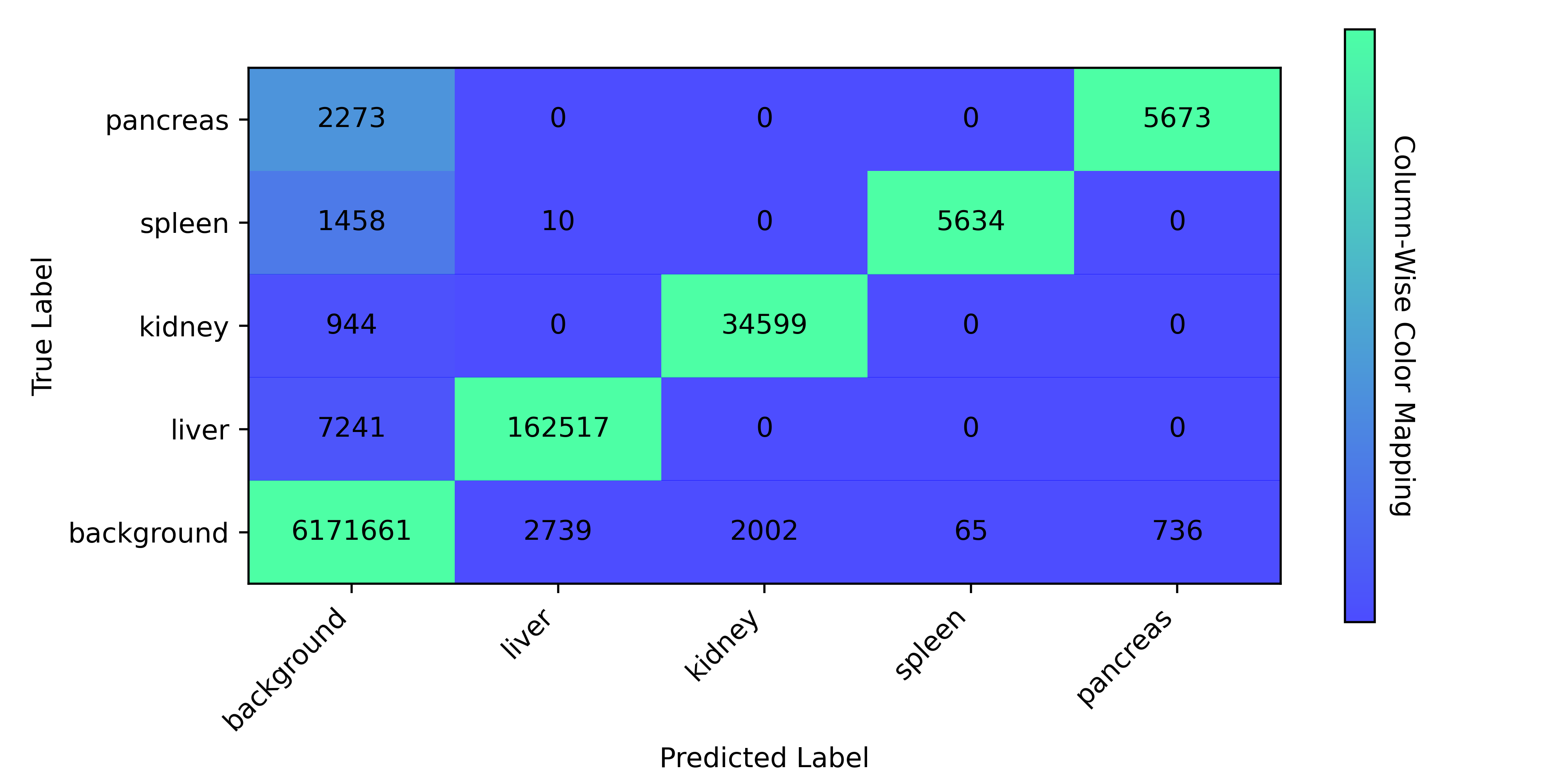}
        \caption{Auto Window}
        \label{fig:SmallScaleExp_cm_AbdomenCT1K_AutoWin}
    \end{subfigure}
    \caption{Confusion Matrix of the AbdomenCT-1K dataset using different window setting method. The units are “voxel.” Following the adoption of Auto Window, a marginal reduction in accuracy is noted for the “liver” category, which had previously demonstrated high performance. Conversely, there is an improvement in accuracy for the anatomically intricate “kidney and pancreas” categories.}
    \label{fig:SmallScaleExp_cm_AbdomenCT1K}
\end{figure}

\begin{figure}[tbp]
    \centering
    \begin{subfigure}[b]{0.32\linewidth}
        \includegraphics[width=\linewidth]{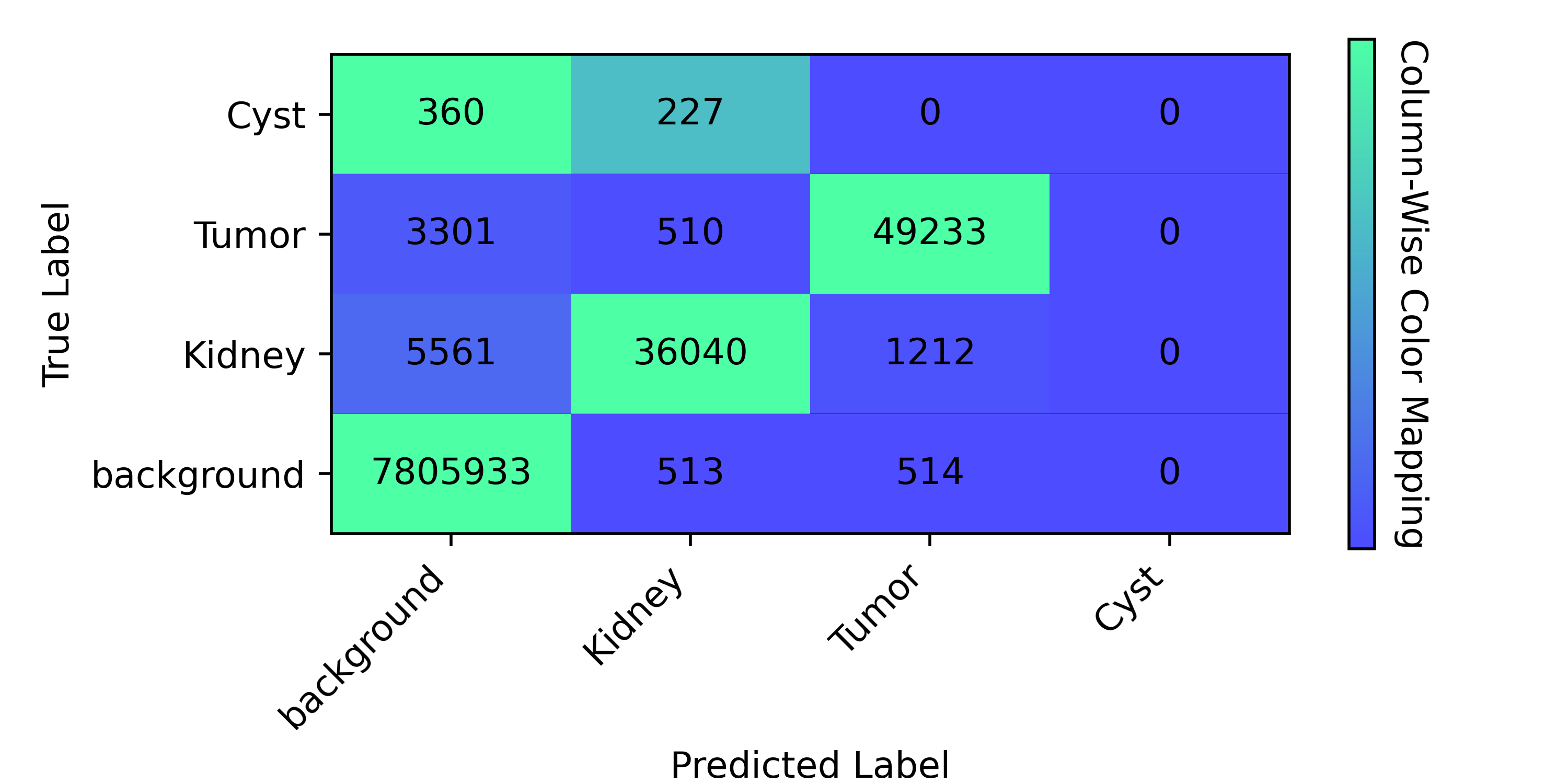}
        \caption{Instance Norm}
        \label{fig:SmallScaleExp_cm_KiTS23_InstNorm}
    \end{subfigure}
    \begin{subfigure}[b]{0.32\linewidth}
        \includegraphics[width=\linewidth]{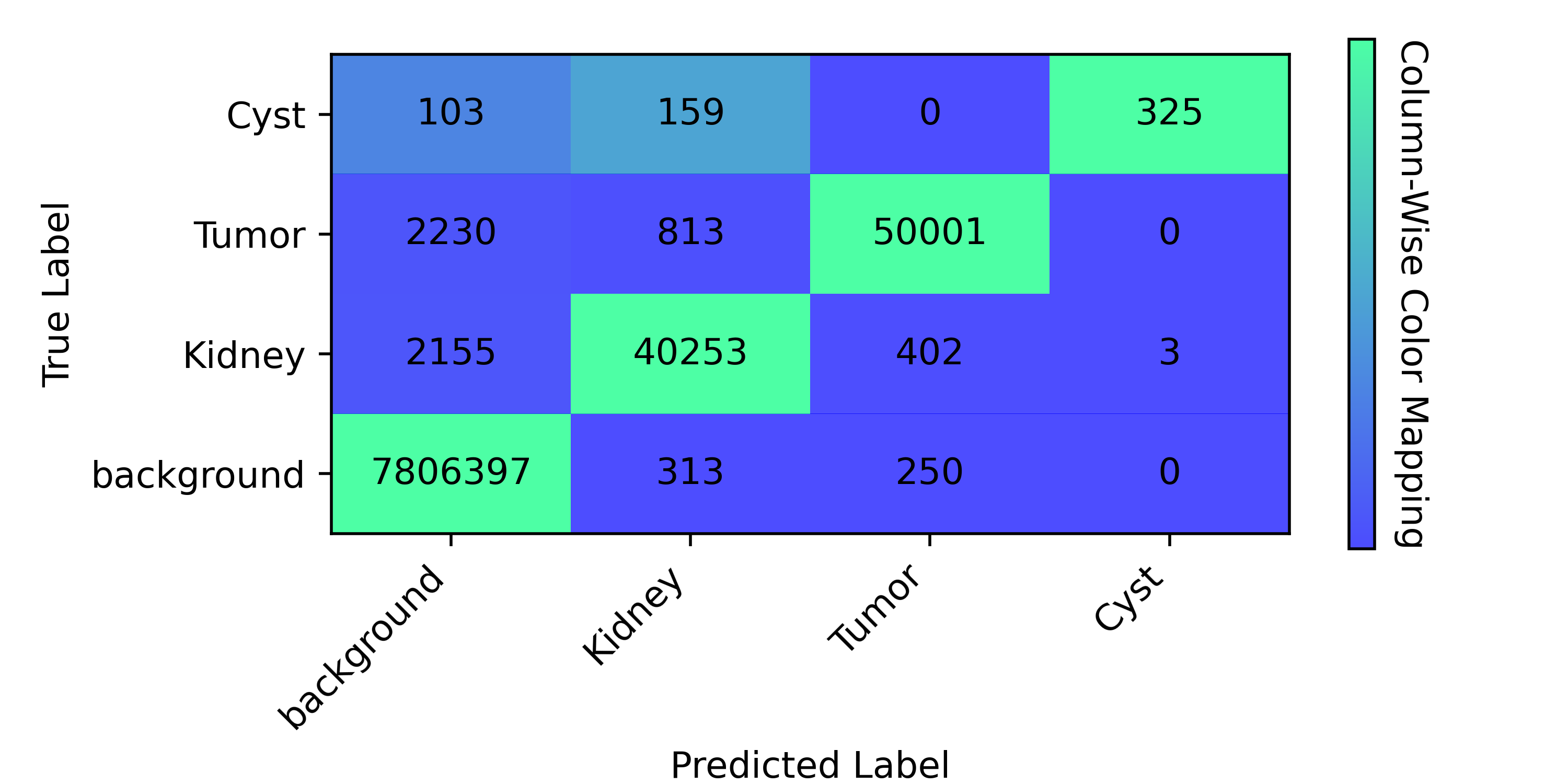}
        \caption{Manual Window}
        \label{fig:SmallScaleExp_cm_KiTS23_ManualWin}
    \end{subfigure}
    \begin{subfigure}[b]{0.32\linewidth}
        \includegraphics[width=\linewidth]{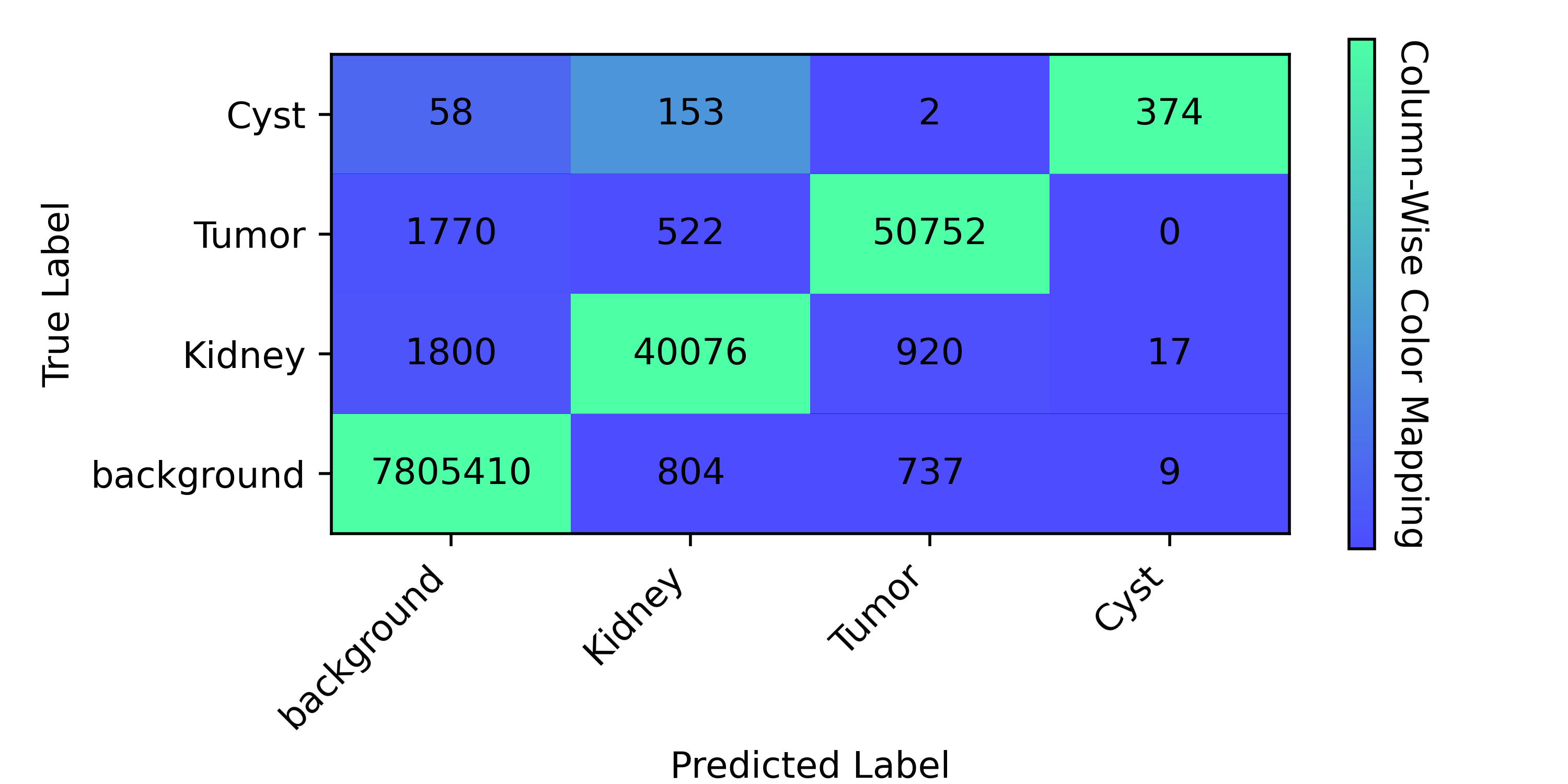}
        \caption{Auto Window}
        \label{fig:SmallScaleExp_cm_KiTS23_AutoWin}
    \end{subfigure}
    \caption{Similar Matrix with \cref{fig:SmallScaleExp_cm_AbdomenCT1K} on KiTS23 Dataset, the Auto Window improves the accuracy on hard class, e.g. Cyst. The accuracy it can achieve even slightly surpasses that of manual windowing.}
    \label{fig:SmallScaleExp_cm_KiTS23}
\end{figure}

The results are shown in \cref{tab:SmallScaleExp}. The advantages of our proposed method are not as obvious in small datasets and small tasks, and its performance on the CT-ORG and KiTS23 datasets is slightly worse than that of manually set CT windows. But the confusion matrixs (\cref{fig:SmallScaleExp_cm_AbdomenCT1K,fig:SmallScaleExp_cm_KiTS23}) indicate that the proposed Auto Window still slightly improves accuracy on several classes which is hard to predict or has fewer areas in annotations. The methodology we have proposed is specifically engineered for complex, large-scale tasks that extend across multiple CT window levels. It is important to note that the effectiveness of our approach may not be fully appreciated in smaller tasks. Considering the broad spectrum of downstream applications in the medical field, our method exhibits robust performance, even under suboptimal scenarios, which suggests that the acceptable robustness of the Auto Window.
\begin{table}[htbp]
    \raggedright
    \caption{Small-Scale dataset performance overview. The proposed Auto Window is robust enough to achieve the manual window's accuracy. } 
    \label{tab:SmallScaleExp}

    \begin{tabular}{rcc|lllllllll}
        \hline
        \multirow[b]{2}{*}{Dataset} & \multirow[b]{2}{*}{Organ} & \multirow[b]{2}{*}{Disease} & \multicolumn{3}{c}{Instance Norm} & \multicolumn{3}{c}{Manual Window} & \multicolumn{3}{c}{Auto Window}\\
        & & & Dice & Recall & Prec. & Dice & Recall & Prec. & Dice & Recall & Prec.\\
        \hline
        AbdomenCT-1K & \checkmark & & 80.61 & 76.11 & 88.74 & 95.32 & \textbf{95.95} & 94.71 & \textbf{95.42} & 94.84 & \textbf{96.02} \\
        CT-ORG    & \checkmark & & 66.80 & 59.76 & 83.19 & \textbf{95.10} & \textbf{95.30} & \textbf{94.92} & 91.34 & 90.99 & 91.75\\
        KiTS23    & \checkmark & \checkmark & 66.40 & 60.36 & 87.76 & \textbf{90.31} & \textbf{93.83} & 87.54 & 89.47 & 86.12 & \textbf{93.76}\\
        ImageTBAD &  & \checkmark & 69.18 & 65.94 & 81.27 & 86.23 & \textbf{94.97} & 81.37 & \textbf{94.54} & 93.22 & \textbf{96.09}\\
        \hline
    \end{tabular}
\end{table}

\subsubsection{Private Dataset}

We have collected a series of high-quality standardized DICOM imaging sequences in clinical scenarios. These patients came from various parts of China to Shanghai for consultation and were diagnosed with gastric cancer. It should be noted that although the common feature of these cases is the presence of gastric cancer lesions, this experiment does not aim at the segmentation of gastric cancer lesion but only attempts to perform general organ and tissue segmentation. In any experiment of this study, the data we collected did not participate in the training. Therefore, the reasoning performance demonstrated in this chapter can verify whether our method can be effective on standardized imaging files, which is also considered part of its clinical potential.

\begin{figure}
    \centering
    \begin{subfigure}[b]{0.45\linewidth}
        \includegraphics[width=\linewidth]{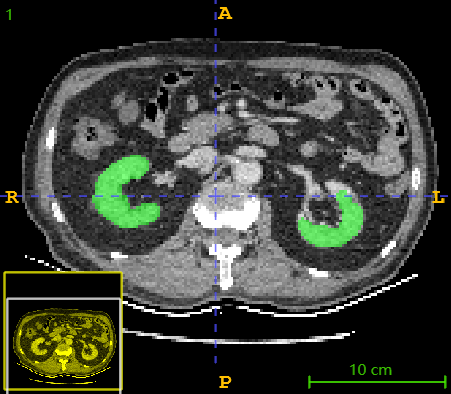}
        \caption{Auto Window - Kidney}
        \label{fig:Pri_AWin8_Kidney}
    \end{subfigure}
    \begin{subfigure}[b]{0.45\linewidth}
        \includegraphics[width=\linewidth]{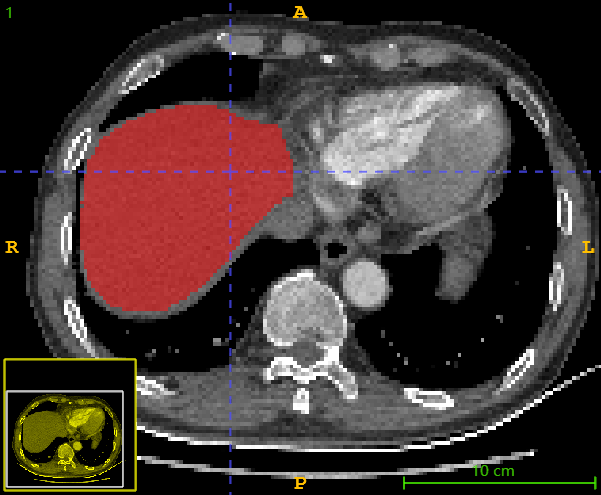}
        \caption{Auto Window - Liver}
        \label{fig:Pri_AWin8_Liver}
    \end{subfigure}
    \caption{Segmentation on private dataset using AbdomenCT1K-Trained model. The results show that when using Auto Window, the model is still able to generate valid predictions on datasets it has never seen before.}
\end{figure}

\subsection{Interpretability}
\label{sec:Exp_Interp}

The design of Auto Window, which decoupling from neural network, enables us to examine the status of the Auto Window at critical mapping points and correlate these with clinical practice. Convergence of the learned window with the empirically derived rules indicates an alignment of knowledge acquired by our method with that of radiologists.

\subsubsection{Window Extractor}

\begin{figure}[tbp]
    \centering
    \begin{subfigure}[b]{\linewidth}
        \includegraphics[width=\linewidth]{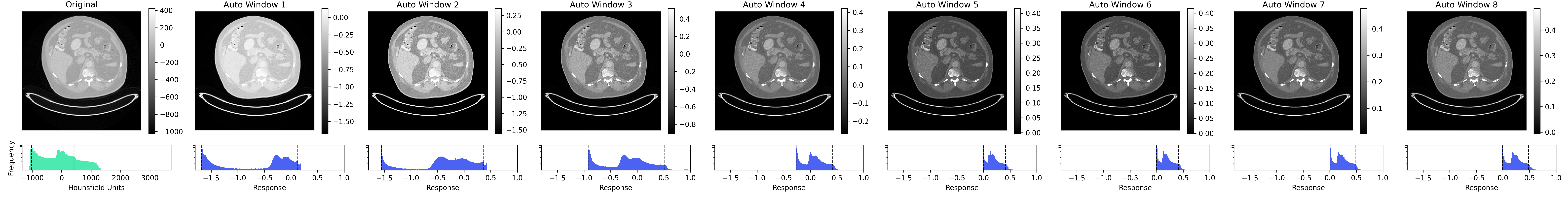}
        \caption{8 Auto Windows}
    \label{fig:WinE_HU_Response_8}
    \end{subfigure}

    \vspace{1em}

    \begin{subfigure}[b]{\linewidth}
        \includegraphics[width=\linewidth]{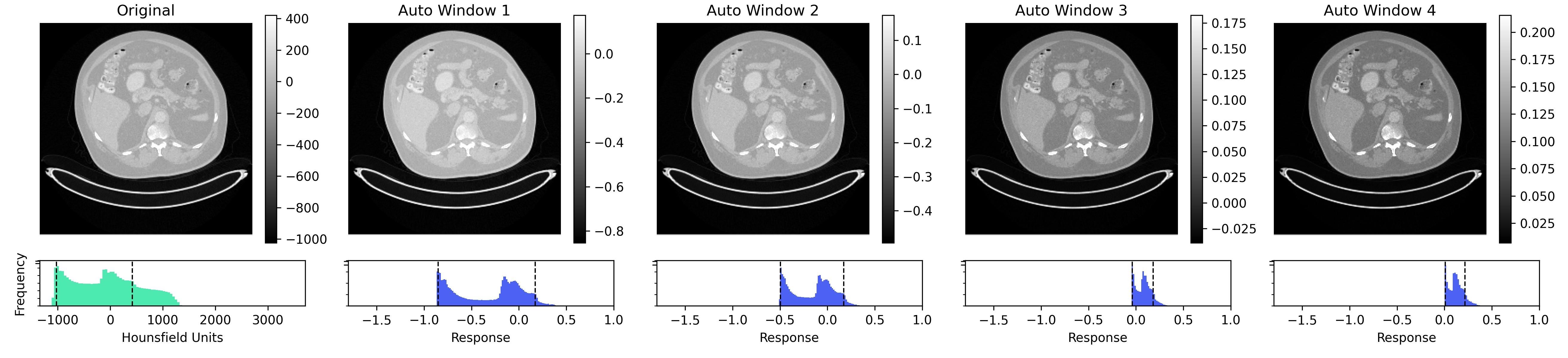}
        \caption{4 Auto Windows}
    \label{fig:WinE_HU_Response_4}
    \end{subfigure}
    \caption{The response of the Window Extractor to the HU value for AbdomenCT1K dataset. Each slice sub-figure illustrates a segment of the HU subdomain, with the precise window delineated by a dashed line in the histogram below. The average values of the HU subdomains that different window extractors focus are trending upwards. The the latter Window Extractors tend to concentrate on comparable HU subdomains.}
    \label{fig:WinE_HU_Response}
\end{figure}

\begin{figure}[tbp]
    \centering
    \begin{subfigure}[b]{\linewidth}
        \includegraphics[width=\linewidth]{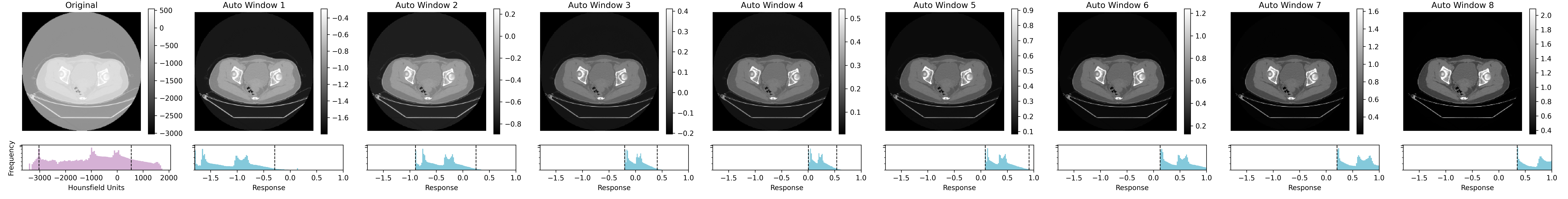}
        \caption{Totalsegmentator Dataset}
    \label{fig:WinE_HU_Response_Tsd}
    \end{subfigure}

    \vspace{1em}

    \begin{subfigure}[b]{\linewidth}
        \includegraphics[width=\linewidth]{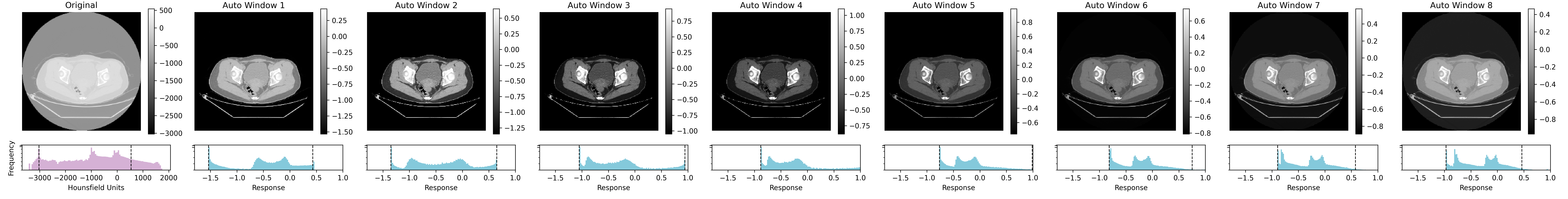}
        \caption{FLARE 2023 Dataset}
    \label{fig:WinE_HU_Response_FLARE2023}
    \end{subfigure}
    \caption{The response on large-scale datasets. Despite the discrepancies in the specific sub-windows across the two datasets, they both incorporate the following types: \textbf{1)} Full window, with most invalid areas automatically removed. \textbf{2)} Medium HU wide window, with rich feature and have a wider response spectrum. \textbf{3)} Medium HU narrow window, with intense feature and a narrower response spectrum. \textbf{4)} High HU narrow window, with bone features highlighted.}
    \label{fig:WinE_HU_Response_LargeDataset}
\end{figure}

Initially, we examine the responses of the Window Extractor for each integer within the range of $[-1024, 3072]$. This analysis provides insights into the modules’ response characteristics to the input HU values, thereby illustrating the Auto Window established by our approach (\cref{fig:WinE_HU_Response}). The Window Extractors demonstrate a marked divergence in their emphasis on HU subdomains when trained with large datasets (\cref{fig:WinE_HU_Response_LargeDataset}). 

\begin{figure}[p]
    \centering
    \begin{subfigure}[b]{\linewidth}
        \centering
        \includegraphics[width=0.9\linewidth]{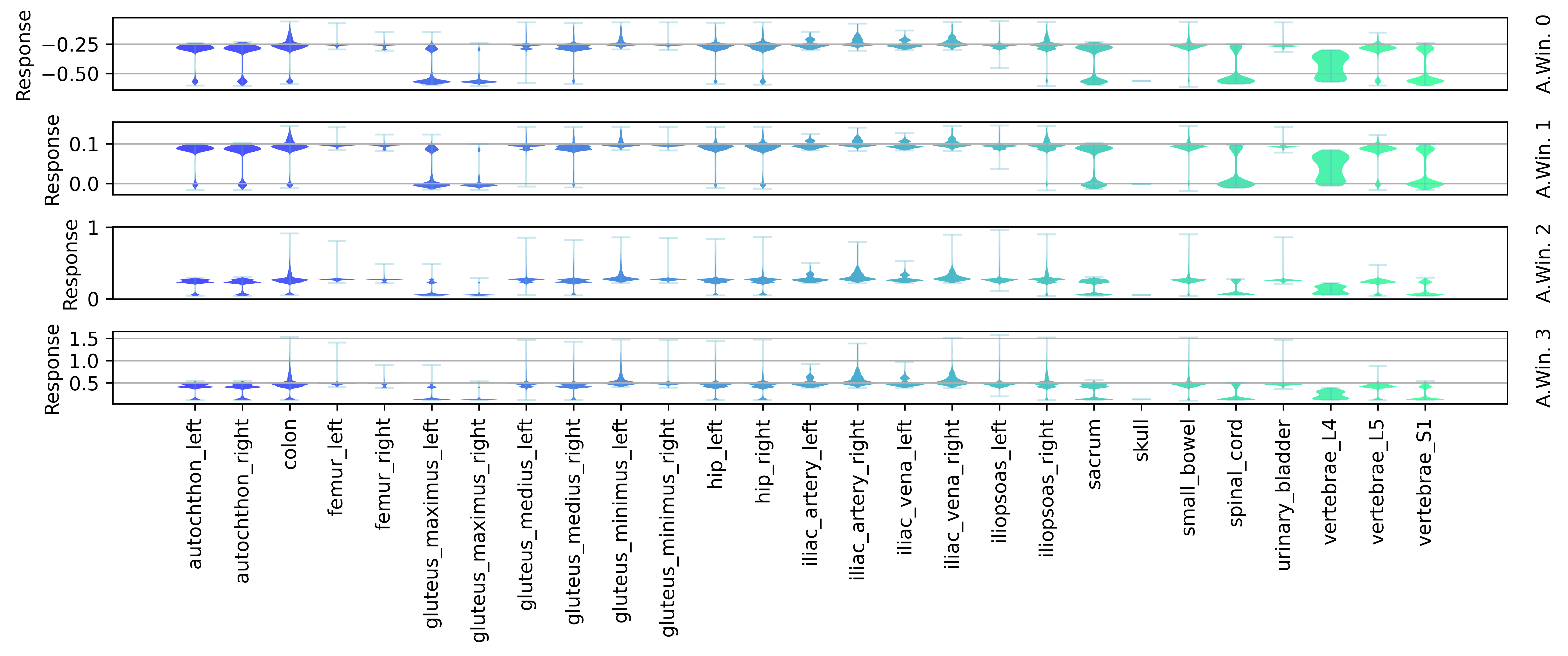}
        \caption{4 Auto Windows}
        \label{fig:HU_violin_4}
    \end{subfigure}
    \begin{subfigure}[b]{\linewidth}
        \centering
        \includegraphics[width=0.9\linewidth]{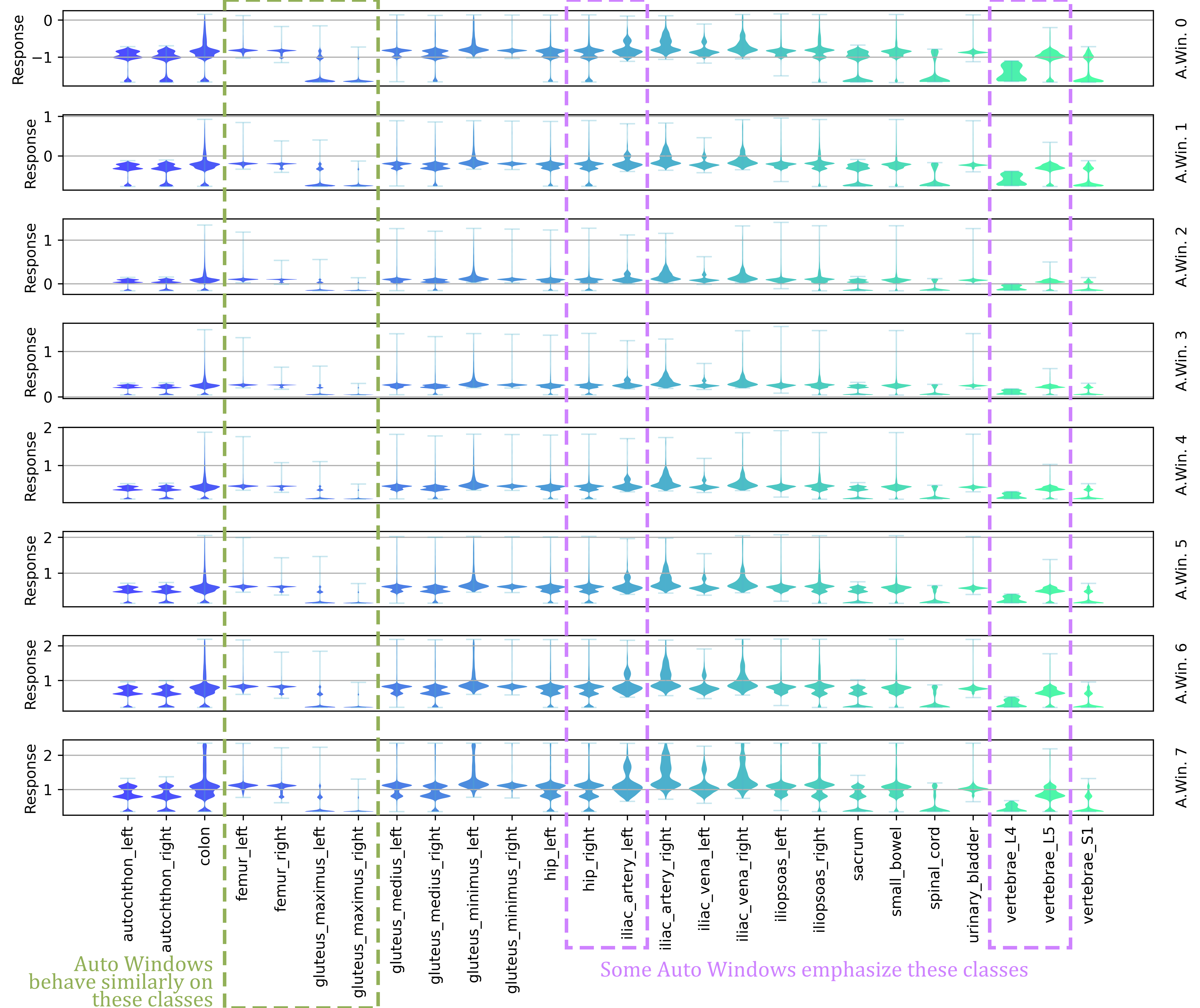}
        \caption{8 Auto Windows}
        \label{fig:HU_violin_8}
    \end{subfigure}
    \caption{The distribution of HU values within the regions predicted by the model for each class. The $X$ axis ticks are labels of Totalsegmentator dataset, and $Y$ axis are dimensionless window response intensity. This illustration enables us to examine the distribution of features contributed by various windows to a single class. The results show that some Auto Windows do indeed show a stronger focus on certain specific classes, but there are also some classes of targets where different Auto Windows have the similar sensitivity.}
    \label{fig:HU_violin}
\end{figure}

In order to validate the aforementioned hypothesis, we identify the regions corresponding to each class predicted by the model’s output and analyze the distribution of HU values within these regions across all sub-windows. This method of visualization enables us to assess the extent of feature richness contributed by each sub-window for the model’s predictions of each class. \cref{fig:HU_violin} visualizes the results from the Totalsegmentator dataset. When using 8 Auto Windows, bone tissue types such as sacrum, vertebrae, and spinal cord (with higher HU values and occuping smaller areas) show more significant differences in responses across different windows. In contrast, muscle and fat types like glutens and hips (with HU values close to 0 and occuping larger sub-domains) exhibit similar HU distribution ranges across different windows. This difference suggests that Auto Windows tend to extract differently on HU subdomains that are more widely distributed, while the extraction for densely distributed HU subdomains is more similar.

\begin{figure}[tbp]
    \centering
    \includegraphics[width=0.5\linewidth]{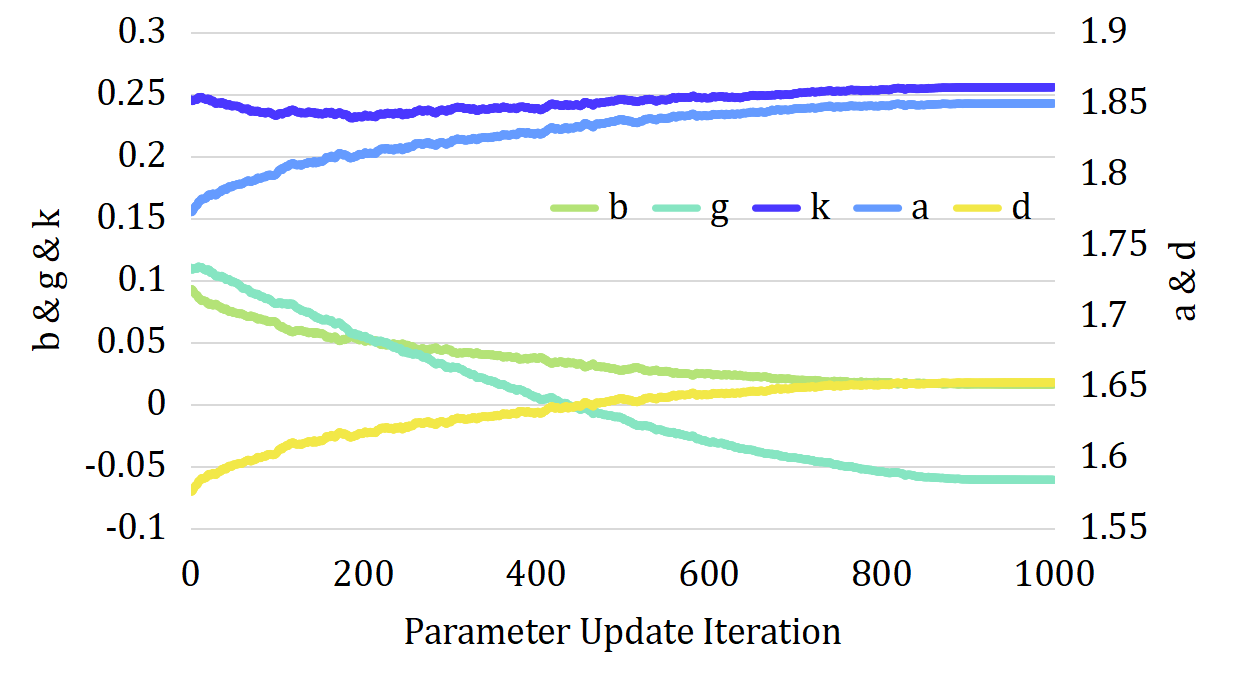}
    \caption{The training process of the Window Extractor during Totalsegmentator training with 4 Auto Windows. These results are from the parameter of the third window. All the parameters that can be learned demonstrate a stable learning process and show a uniform trend of adaptive regulation. In this Window Extractor, the parameters $b$ and $g$ are decreasing, while all the other parameters are increasing. This suggests that the current window has a preference for a lower HU region, and the window’s receptive field is continuously expanding, indicating that the Extractor is gathering information from a wider range of HU domains.}
    \label{fig:WinE_TrainTime}
\end{figure}

Due to the ease of mathematical analysis in the design of the Window Extractor’s learnable parameters, we monitored the variations of these parameters throughout the training process. This enables us to analyze the neural network’s feature extraction at different stages of training. As has been mentioned in \cref{sec:Method_WinE}, there are three major parameters that controls the window location, namely $a, b, d$. Among them, $a$ and $b$ are the fine-tuning parameters, and $d$ is the large-scale control parameter. So $d$ will be the most important parameter to monitor when assessing window location adjustment process during training. The results are shown in \cref{fig:WinE_TrainTime}. Throughout the training depicted in this figure, the relevant window consistently shifts towards lower HU levels, with a slight preference for increased width. In the final stages of training, the implementation of a reduced learning rate protocol results in the stabilization of all parameters. The remaining auto windows demonstrate a comparable trend of stability. In light of the potential for significant fluctuations in the upper network due to unstable outputs from any underlying neurons, the method proposed by us can be deemed to possess adequate robustness.

\subsubsection{Tanh Rectifier}

\begin{figure}[tbp]
    \centering
    \includegraphics[width=0.24\linewidth]{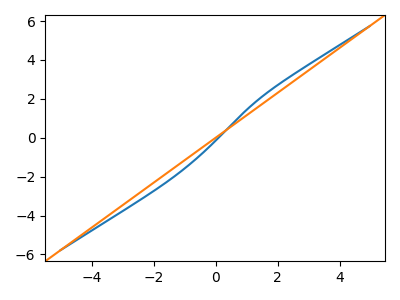}
    \includegraphics[width=0.24\linewidth]{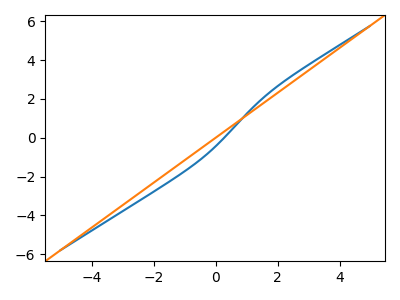}
    \includegraphics[width=0.24\linewidth]{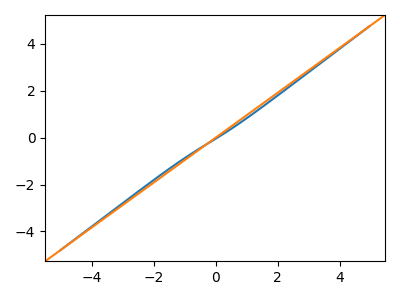}
    \includegraphics[width=0.24\linewidth]{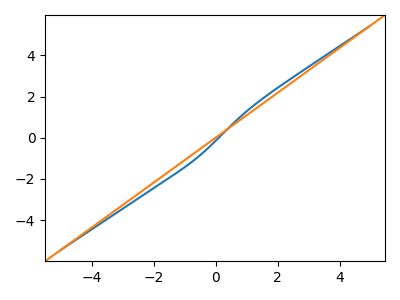}
    \caption{The learned Tanh-Rectifiers' response during Totalsegmentator training with 4 Auto Windows. The $x$-axis shows the input signal, while the $y$-axis displays the output response. All four Tanh-Rectifier combinations exhibit a mild degree of signal modification, with adjustments predominantly occurring in the vicinity of the origin.}
    \label{fig:Tanh_HU_Response}
\end{figure}

Each Window Extractor is complemented by a Tanh Rectifier for the fine-tuning of the extracted values (\cref{sec:Method_TRec}). Our design hopes that this module does not perform excessive mappings, as it lacks the high level of explainability provided by the Window Extractor’s adaptive parameters. The overly aggressive corrections would diminish the reliability of the analysis of the Window Extractor’s parameters, as the Window Extractor would no longer accurately represent the method’s overall automatic adjustment of the window. We evaluated the response characteristics of the Tanh Rectifier using the same input values, i.e. $[-1024, 3072]$. The findings indicate that the module’s response curve closely resembles $y = x$, indicating a mapping that are close to identity, with only minimal nonlinear variations in a few narrow window settings (\cref{fig:Tanh_HU_Response}).

Indeed, deciphering the learning characteristics of the tanh rectifier poses a challenge. When trained on multi paralleled window and extensive datasets, the subtle adjustments within narrow window ranges influence all categories. The distinction between certain classes, such as the pancreas and duodenum, is ambiguous due to their proximity in HU values. Consequently, the neural network frequently encounters difficulties in determining an optimal HU value for differentiating these classes. In such scenarios, deploying a tanh rectifier that specifically targets narrow windows could potentially enhance the delineation of these boundaries, thereby improving the segmentation performance for specific classes. Based on our analysis results above, the tanh rectifier is not recognized as the primary functional module in the method we have proposed.

\subsubsection{Paralleled Windows Fusion}

The fusion mechanism, positioned as the sub-component nearest to the output of the proposed method, employs a low-dimensional, adjustable weight matrix to dictate how each window extracts the relevant signals from all the others. Before commencing this step, the extraction of features by all sub-windows is conducted independently and in parallel. There are no established channels for information exchange among them. Instead, they sequentially and linearly allocate their initial regions of interest within the HU sub-domain. Moreover, our design approach is geared towards ensuring that each sub-window focuses on distinctive features. Hence, we propose that even in the context of Auto-Window-Wise fusion, there should be constraints on the distribution of fusion weights to prevent it from being overly aggressive.

\begin{figure}[tbp]
    \centering
    \begin{subfigure}{0.45\linewidth}
        \includegraphics[width=\linewidth]{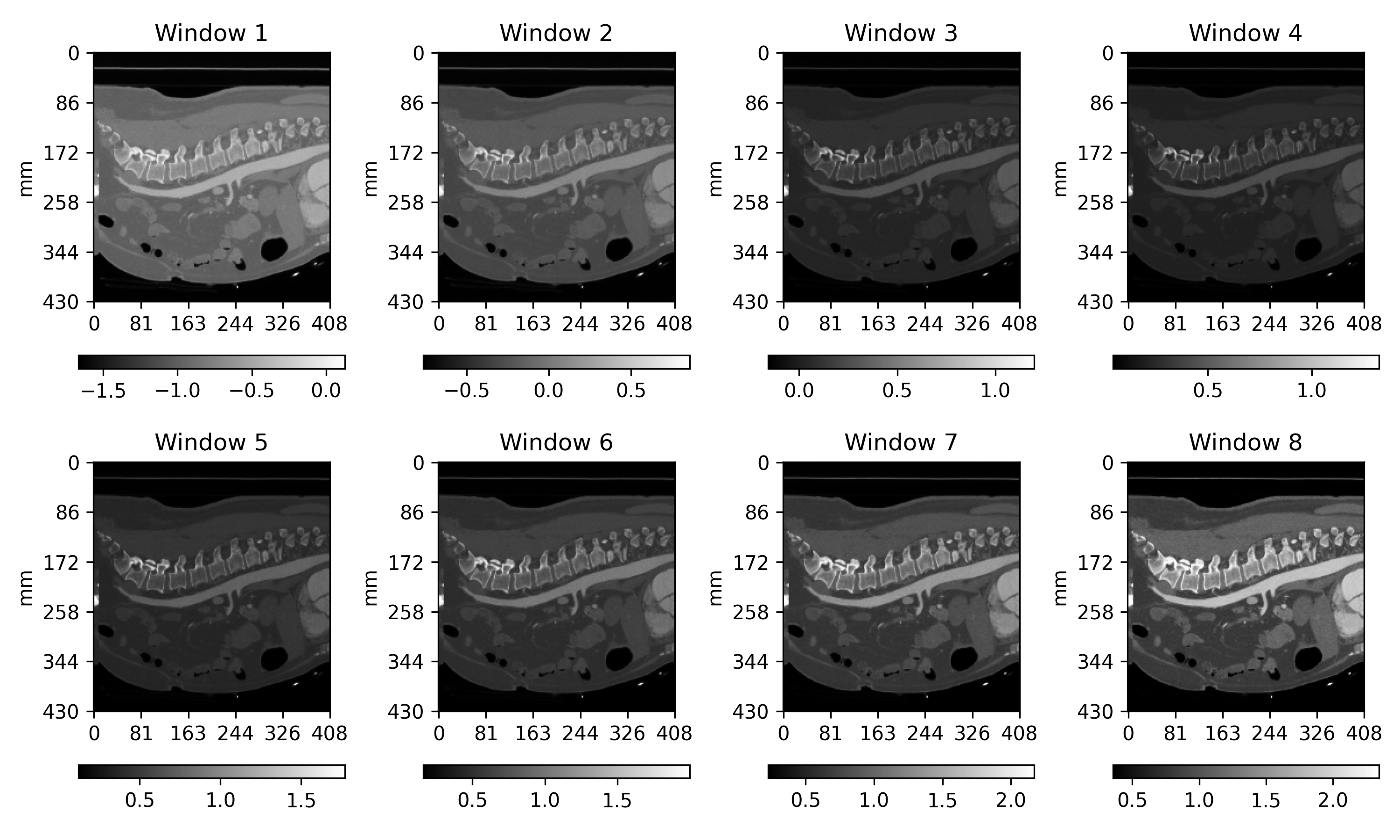}
        \vspace{1em}
        \includegraphics[width=\linewidth]{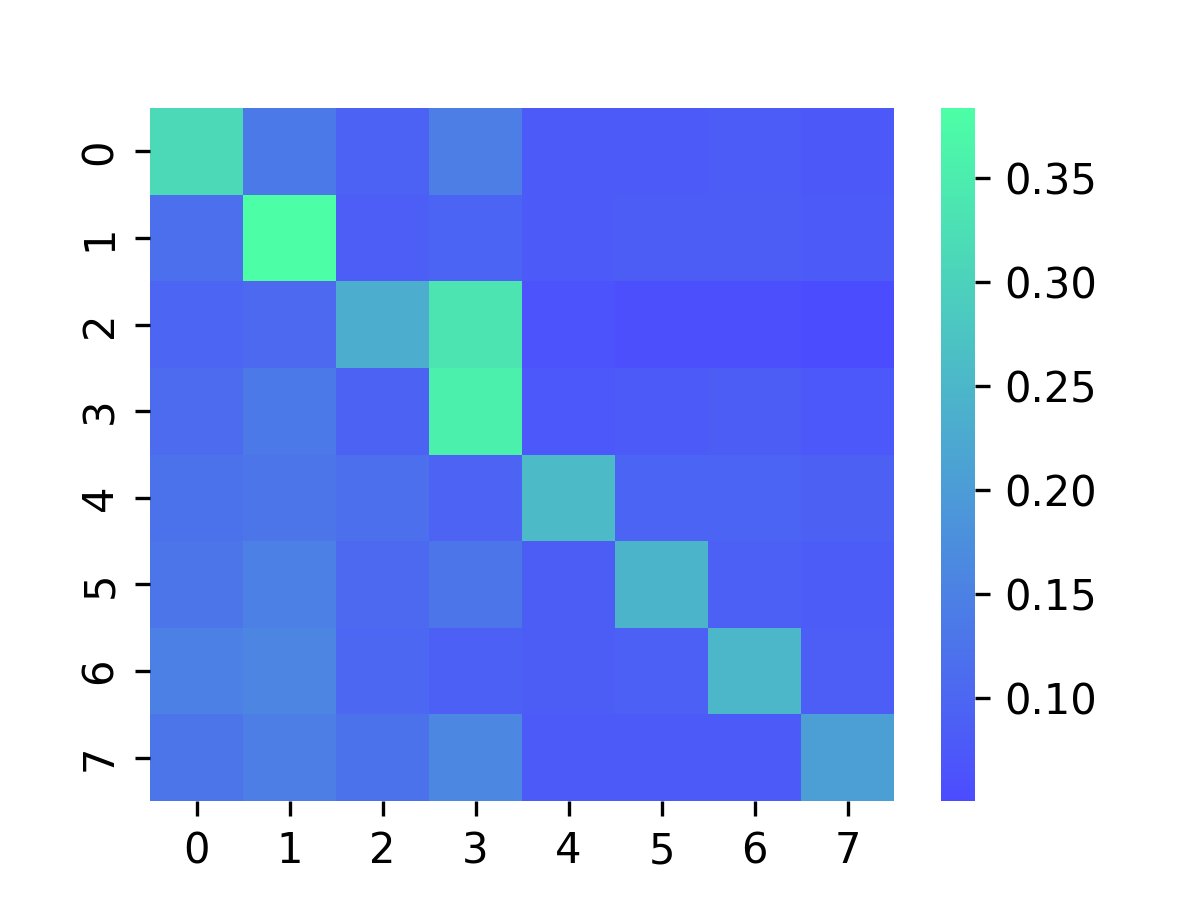}
        \caption{Totalsegmentator}
    \end{subfigure}
    \hspace{3em}
    \begin{subfigure}{0.45\linewidth}
        \includegraphics[width=\linewidth]{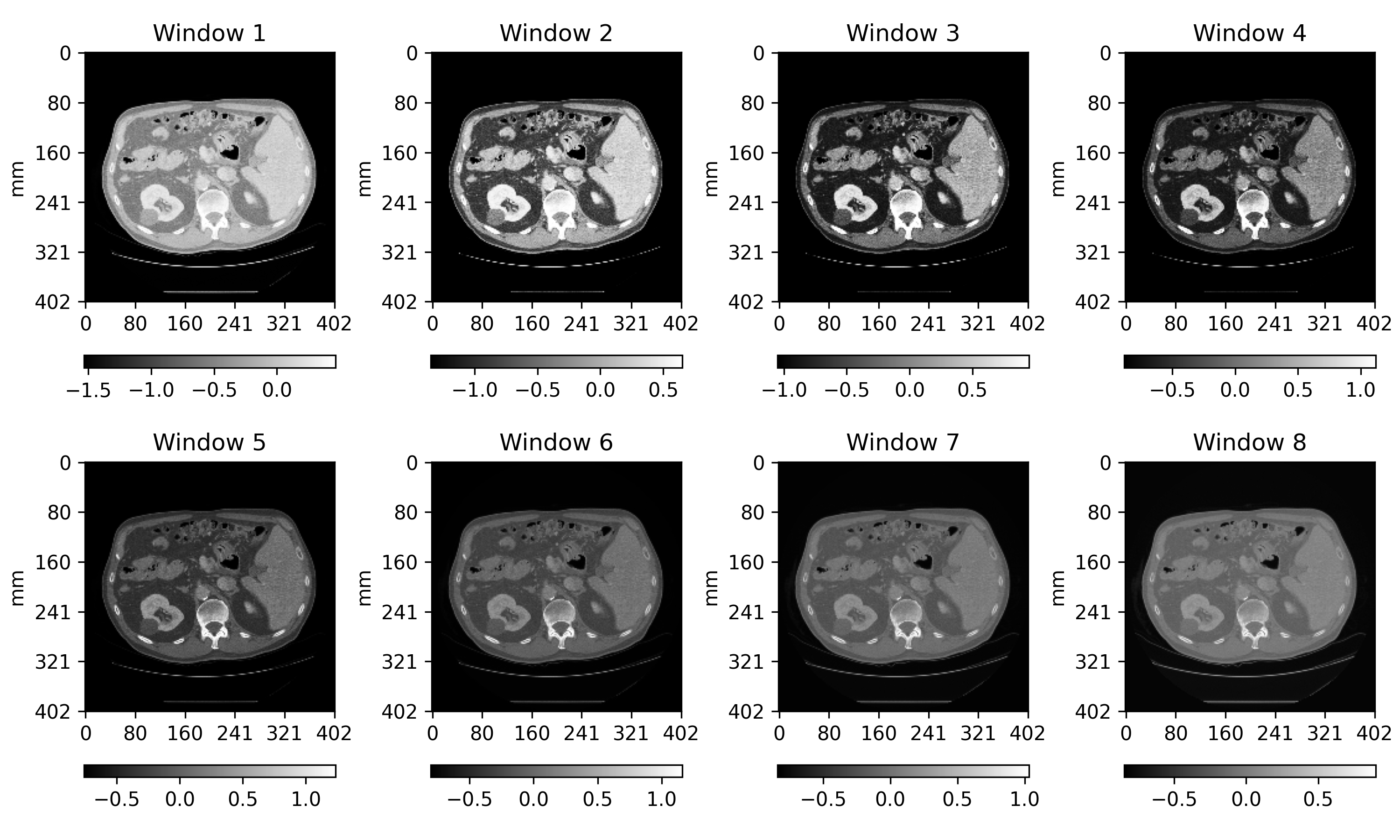}
        \vspace{1em}
        \includegraphics[width=\linewidth]{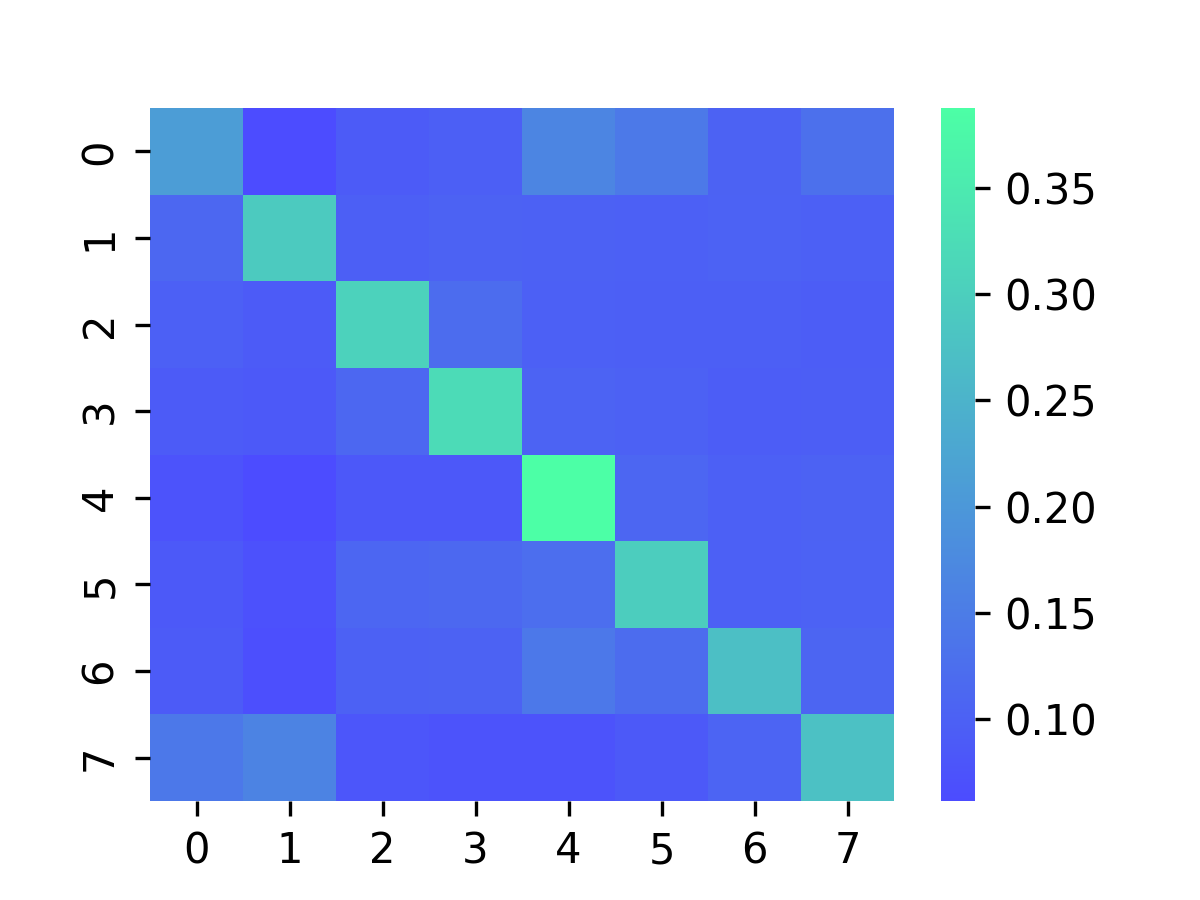}
        \caption{FLARE 2023}
    \end{subfigure}
    \caption{The weight matrices of the Paralleled Window Fusion module (2nd. row) along with each sub-auto-window's extraction output (1st. row). For the weight matrix, the vertical axis represents the fused sub-window channels, while the horizontal axis represents the channels before fusion. When training on different datasets, the automatic window extraction strategy varies, the weight matrix varies as well.}
    \label{fig:CrsF}
\end{figure}

Based on the considerations above, the initial configuration of the weight matrix is set to an identity matrix, indicating that it is designed to focus exclusively on its own signals without extracting any information from other modules. Post-training analysis (\cref{fig:CrsF}) reveals that the weight matrices retain their diagonal pattern while exhibiting nuanced extraction favoritism. Upon detailed analysis, we found few consistent learning features among the obtained multiple weight matrices. When using 8 automatic windows, the weight matrix preferred Auto Window No.0,1,3 when running on Totalsegmentator, but preferred windows Auto Window No.4,5 when running on FLARE 2023. Considering that the module under analysis is impacted by the neural network, the origin of this discrepancy remains undetermined. Nonetheless, it is plausible that it correlates with the distinct distribution of HU subdomains within the recognition targets.

\subsection{Performance Overhead}
\label{sec:Exp_Perf}

The proposed method almost does not increase the additional computational space requirements, with each Window Extractor having 5 learnable parameters, each Tanh having $2\kappa$ learnable parameters, and Cross-Window Fusion having a weight matrix of size $N \times N$. Compared to the millions of parameters in neural networks, the number of parameters in this method is negligible.

\section{Discussion and Conclusion}
\label{sec:Conc}

In this study, we propose an Auto CT Window Setting module that serves as a precursor for the majority of current neural networks. This module assists neural networks in extracting diverse features from different windows in CT data with high dynamic range, thereby enhancing the performance of large-scale medical segmentation tasks with multiple objectives. We have witnessed encouraging performance gains on several large datasets, suggesting that our proposed method may help reduce the complexity of constructing medical segmentation tasks based on deep learning, and can deliver superior performance in complex tasks.

Simultaneously, the performance boost on smaller datasets is modest, indicating that in more specialized tasks, the proposed Auto Window does not significantly outperform the Window set by empirical rules. In disease-oriented medical AI, such tasks are quite important. A more robust window adaptation algorithm within narrower windows might achieve better performance in such applications.

The automatic windowing mechanism can be conceptualized as an adaptive remapping of the HU distribution. The development of more sophisticated adaptive methods may allow for the automatic mapping of scanning sequences from arbitrary domains into a specified, limited latent space. Such advancements could render the same neural network weights applicable across datasets generated by diverse scanning protocols, potentially reducing the costs and complexities associated with CT sequence analysis. This approach bears resemblance to registration techniques; however, it offers greater generality as it does not necessarily require a precise definition of the input signal, unlike traditional registration methods.

\printcredits

\bibliographystyle{cas-model2-names}
\bibliography{AutoWindow.bib}

\appendix

\section{Implementation Details}

All experiments were conducted on two devices: one equipped with 4 RTX 4090 GPUs and the other with a RTX 4070 Ti and a RTX 2070 Super GPU. The former is utilized for training on large datasets, while the latter is dedicated to training on small datasets. To ensure consistency and avoid performance variations due to differences in neural network operations, all experiments are conducted using PyTorch 2.5.0, MMEngine 0.10.5, and CUDNN 9.1.0. The detailed MMEngine-Style configurations are available in our github repo.

\end{document}